\documentclass[aps, prd, reprint, amsmath, amssymb, superscriptaddress, floatfix, nofootinbib]{revtex4-1}

\usepackage[pdftex]{graphicx}
\DeclareGraphicsExtensions{.jpg,.png,.pdf}

\usepackage[utf8]{inputenc}
\usepackage{hyperref}
\usepackage{amsmath}
\usepackage{mathrsfs}
\usepackage{amssymb}
\usepackage{amsfonts}
\usepackage{tabularx}
\usepackage{booktabs}
\usepackage{graphicx}
\usepackage{color}
\usepackage{multirow}
\usepackage[perpage,symbol]{footmisc}
\usepackage{verbatim}
\usepackage{dcolumn}
\usepackage{bm}
\usepackage[inline]{enumitem}
\graphicspath{{Fig/}}
\definecolor{theblue}{RGB}{0,50,230}
\hypersetup{
  colorlinks=true,
  linkcolor=theblue,
  citecolor=theblue,
  urlcolor=theblue
}

\newcommand{\pt}{\ensuremath{p}_{T}}

\begin{document}

\title{Unraveling collisional energy loss of a heavy quark in quark-gluon plasma}

\author{Jiazhen~Peng}
\affiliation{%
College of Science, China Three Gorges University, Yichang 443002, China\\
}%
\affiliation{%
Center for Astronomy and Space Sciences, China Three Gorges University, Yichang 443002, China\\
}%

\author{Kewei~Yu}
\affiliation{%
College of Science, China Three Gorges University, Yichang 443002, China\\
}%
\affiliation{%
Center for Astronomy and Space Sciences, China Three Gorges University, Yichang 443002, China\\
}%

\author{Shuang~Li}
\email{lish@ctgu.edu.cn}
\affiliation{%
College of Science, China Three Gorges University, Yichang 443002, China\\
}%
\affiliation{%
Center for Astronomy and Space Sciences, China Three Gorges University, Yichang 443002, China\\
}%

\author{Wei~Xiong}
\email{xiongw@ctgu.edu.cn}
\affiliation{%
College of Science, China Three Gorges University, Yichang 443002, China\\
}%
\affiliation{%
Center for Astronomy and Space Sciences, China Three Gorges University, Yichang 443002, China\\
}%
\author{Fei~Sun}
\email{sunfei@ctgu.edu.cn}
\affiliation{%
	College of Science, China Three Gorges University, Yichang 443002, China\\
}%
\affiliation{%
	Center for Astronomy and Space Sciences, China Three Gorges University, Yichang 443002, China\\
}%

\author{Wei~Xie}
\email{xiewei@ctgu.edu.cn}
\affiliation{%
College of Science, China Three Gorges University, Yichang 443002, China\\
}%
\affiliation{%
Center for Astronomy and Space Sciences, China Three Gorges University, Yichang 443002, China\\
}%

\date{\today}%

\begin{abstract}
At leading order in QCD coupling constant, we compute the energy loss per traveling distance of a heavy quark $dE/dz$
from elastic scattering off thermal quarks and gluons at a temperature $T$,
including the thermal perturbative description of soft scatterings ($-t<-t^{\ast}$) and
a perturbative QCD-based calculation for hard collisions ($-t>-t^{\ast}$).
Within this soft-hard factorization model, we find that the full results of $dE/dz$
behaves a mild sensitivity to the intermediate cutoff $t^{\ast}$,
supporting the validity of the soft-hard approach within the temperature region of interest.
We re-derive the analytic formula for $dE/dz$
in the high-energy approximation, $E_{1}\gg m^{2}_{1}/T$,
where $E_{1}$ is the injected heavy quark energy and $m_{1}$ is its mass.
It is realized that the soft logarithmic contribution,
$dE/dz\propto ln(-t^{\ast}/m^{2}_{D})$,
arises from the $t$-channel scattering off thermal partons,
while the hard logarithmic term, $dE/dz\propto ln[E_{1}T/(-t^{\ast})]$,
stems from the $t$-channel scattering off thermal partons, and
the one $dE/dz\propto ln(E_{1}T/m^{2}_{1})$ comes from the $s$- and $u$-channel scattering off gluons. 
The sum of these contributions cancels the $t^{\ast}$-dependence as observed in the full result.
The mass hierarchy is observed $dE/dz(charm)>dE/dz(bottom)$.
Our full results are crucial for a better description of heavy quark transport in QCD medium,
in particular at low and moderate energy.
We also calculate the energy loss by imposing the Einstein's relationship.
The related results appear to be systematically larger than that without imposing the Einstein's relationship.
\end{abstract}


\maketitle

\section{Introduction}\label{sec:Intro}
The normal nuclear matter turns into a new state of matter characterized by the deconfined partons known as quark-gluon plasma (QGP),
at extremely high temperature and energy density as achieved in the microseconds after the Big Bang~\cite{Gyulassy05, Shuryak05}.
In the past two decades, high energy heavy-ion collisions carried at the Relativistic Heavy-Ion Collider (RHIC)
and the Large Hadron Collider (LHC) provides a unique opportunity for shedding light on the perturbative region of quantum chromodynamics (QCD),
the most understandable of the fundamental interactions in the Standard Model~\cite{Muller12, Shuryak17, Kharzeev:2020jxw}.

Heavy quarks (charm and bottom) are of particular interest probes of the QGP as they are produced
in initial hard scatterings in the early stage of the collision and subsequently propagate through
the QCD medium of quarks, anti-quarks and gluons in thermal equilibrium at a temperature $T$,
resulting in the collisional and radiative energy loss, $dE/dz$,
via elastic and inelastic interactions, respectively~\cite{GYULASSY1994583, HQQGPRapp10, HFQM17Greco, CUJET3QM17}.
This medium-induced effect can be studied using the experimental observables,
such as the relevant production cross-section, nuclear modification factor, elliptic flow and azimuthal correlations.
Thus, the properties of heavy quark energy loss are of intense interest in connection with
the signatures of the formation of QGP in ultra-relativistic heavy-ion
collisions~\cite{RalfSummary16, HFSummaryGROUP16, HFSummaryAarts17, RalfSummary18, He:2022ywp}.

In 1982 Bjorken provided~\cite{EnergyLossMovivateJDB82} a perturbative calculation of
the collisional energy loss of a massless parton due to the elastic scattering off
the thermal quarks and gluons in the QGP.
He estimated $dE/dz$ at leading order in $g$, by making several approximations,
such as (1) assuming an energetic parton, i.e. in the large energy limit;
(2) keeping only the logarithmically divergent integral over momentum transfer;
(3) imposing physically reasonable upper and lower limits to regulate the infrared and ultraviolet divergences.
Finally, it was realized that the collisional energy loss was path-independent and
that it depended on the energy of the parton only logarithmically, see Eq.~\ref{eq:Bjorken} for details.

However, these results suffered from an ambiguity associated with the choice of the upper and low limits for the momentum transfer.
The improvements over the Bjorken approach have been achieved by the subsequent studies,
including more careful treatment of the infrared divergences~\cite{THOMA1991491},
ultraviolet divergences~\cite{PhysRevD.44.1298, PhysRevD.44.R2625},
the inclusion of the running of the coupling~\cite{PhysRevD.77.114017, PhysRevLett.97.212301} and many more.

The aim of this paper is to obtain the heavy quark energy loss $dE/dz$ at low and moderate energy,
by a complete calculation at leading order QCD coupling constant
for the elastic scattering off thermal light quarks and gluons in a QGP.
To separate the contributions from the soft ($\sqrt{-t}\sim gT$) and hard regions ($\sqrt{-t}\gtrsim T$) of the momentum transfer $t$,
an arbitrary momentum scale $t^{\ast}$ is introduced and, subsequently, adjusted according to the comprehensive model-data comparisons.
Concerning the soft component, it is recalculated taking into account the contributions from low-momentum transfer,
and the resulting long-wavelength gluon are screened by the dense mediums.
As a consequence,
the propagator in the gluon-exchange diagrams is replaced by the hard-thermal loop (HTL) propagator~\cite{PhysRevLett.64.1338, Braaten91PRL}.
For the hard component, the hard gluon exchange is considered and the tree-level propagator is used in our calculations~\cite{Li_2021}.

The paper is organized as follows. In Sect.~\ref{sec:eloss}. we focus on the calculation of the
collisional energy loss of a heavy quark crossing a Quark-Gluon Plasma.
In Sect.~\ref{subsec:gamma_HTLpQCD} we introduce the general setup of the employed soft-hard factorization approach,
in particular, the relevant scattering rate, which is crucial for the calculation of the energy loss.
The results in the soft and hard regions are obtained and discussed in Sect.~\ref{subsec:eloss_HTLpQCD}.
Section~\ref{subsec:approximation} is dedicated to the description of the theoretical results in the high-energy limit.
In Sect.~\ref{subsec:eloss_ER} we argue that the collisional energy loss can be directly related to the drag coefficient,
which is one of the three key parameters in the Fokker-Planck and Langevin dynamics.
In Sect.~\ref{sec:result} we show the momentum and temperature dependence of charm and bottom quarks,
as well as the systematic comparisons with other approaches.
Section~\ref{sec:summary} contains the summary and discussion.

\section{Energy loss in the soft-hard factorized approach}\label{sec:eloss}
\begin{figure}[!htbp]
	\centering
	\includegraphics[width=.23\textwidth]{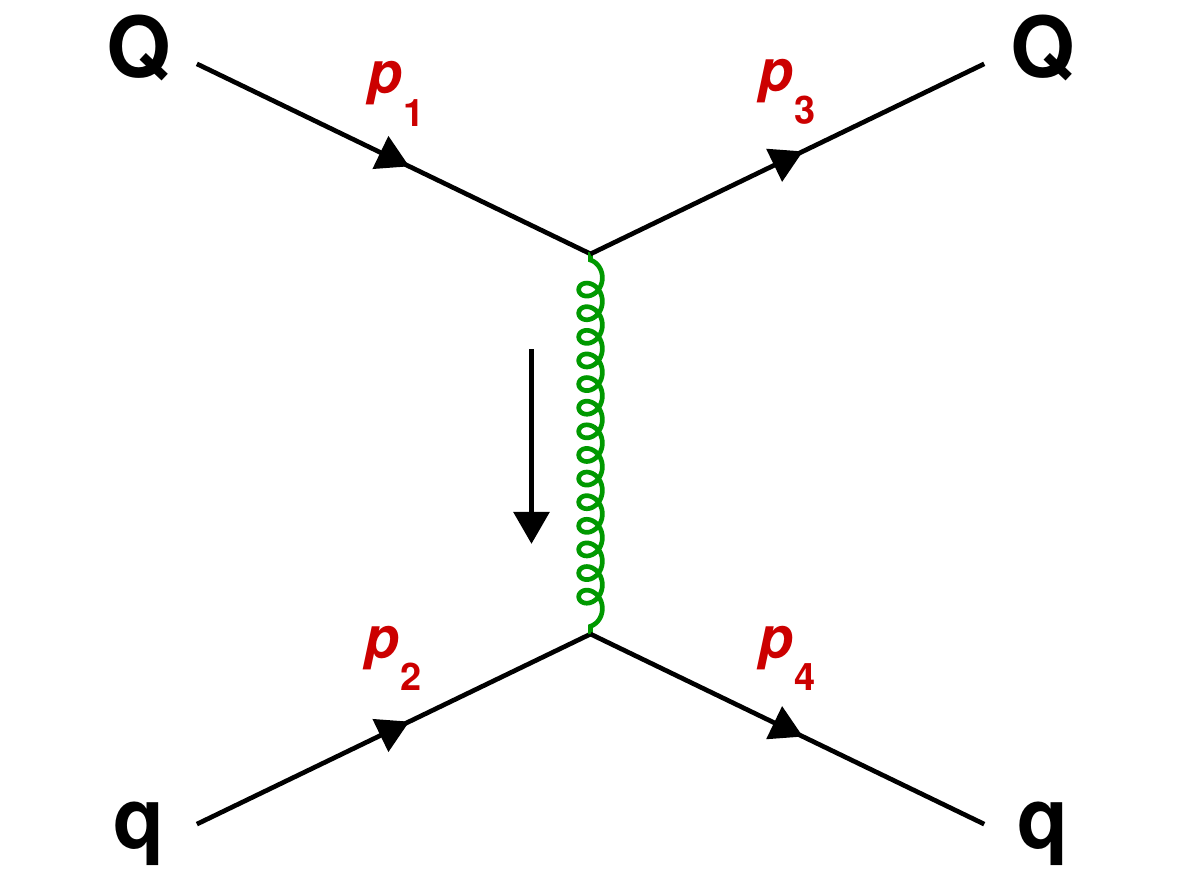}
	\includegraphics[width=.23\textwidth]{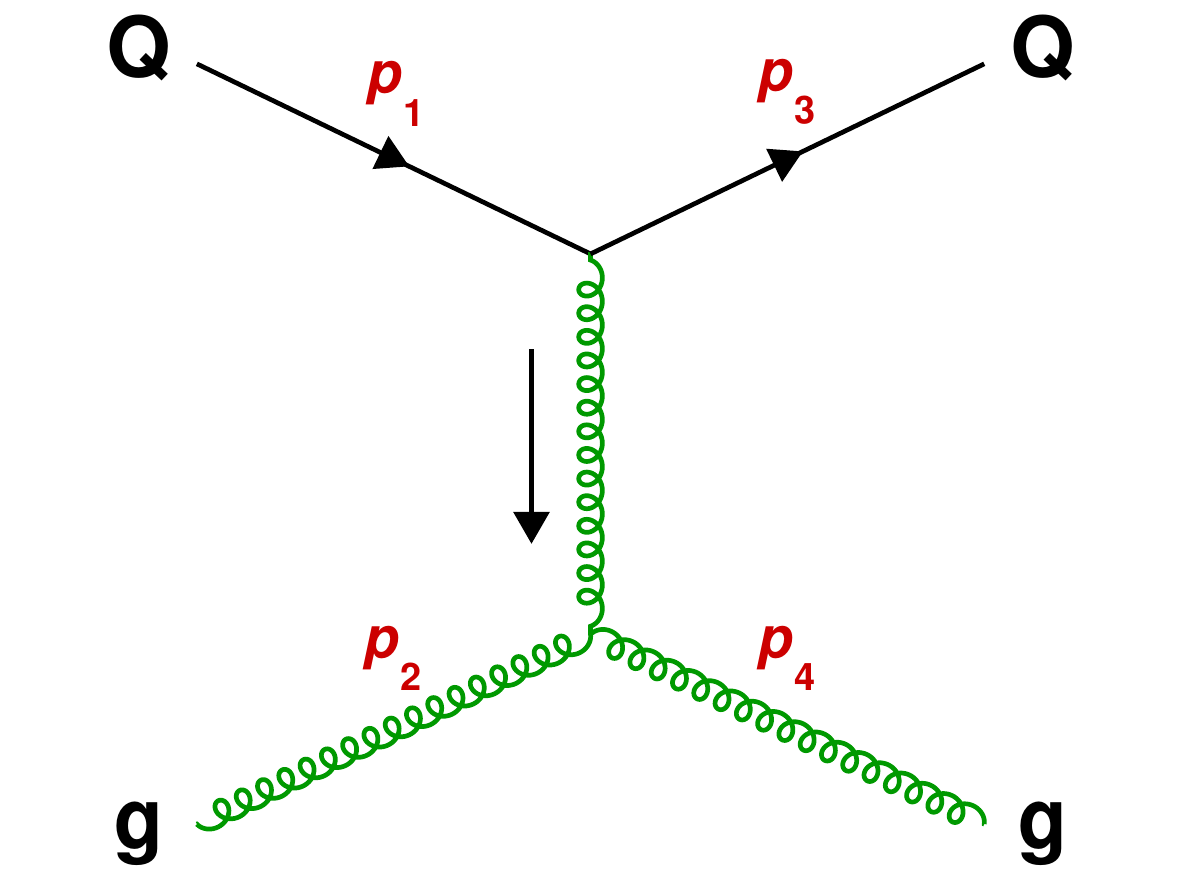}
	\includegraphics[width=.23\textwidth]{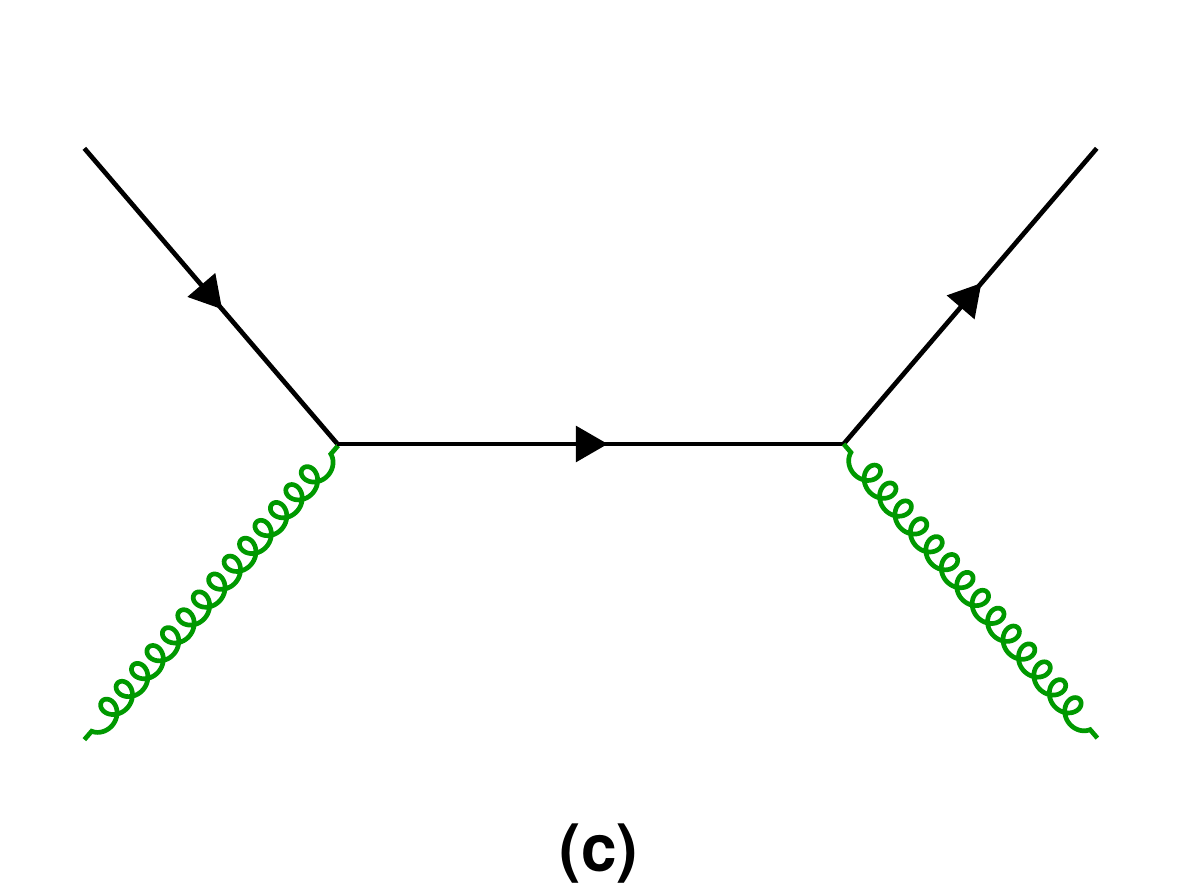}
	\includegraphics[width=.23\textwidth]{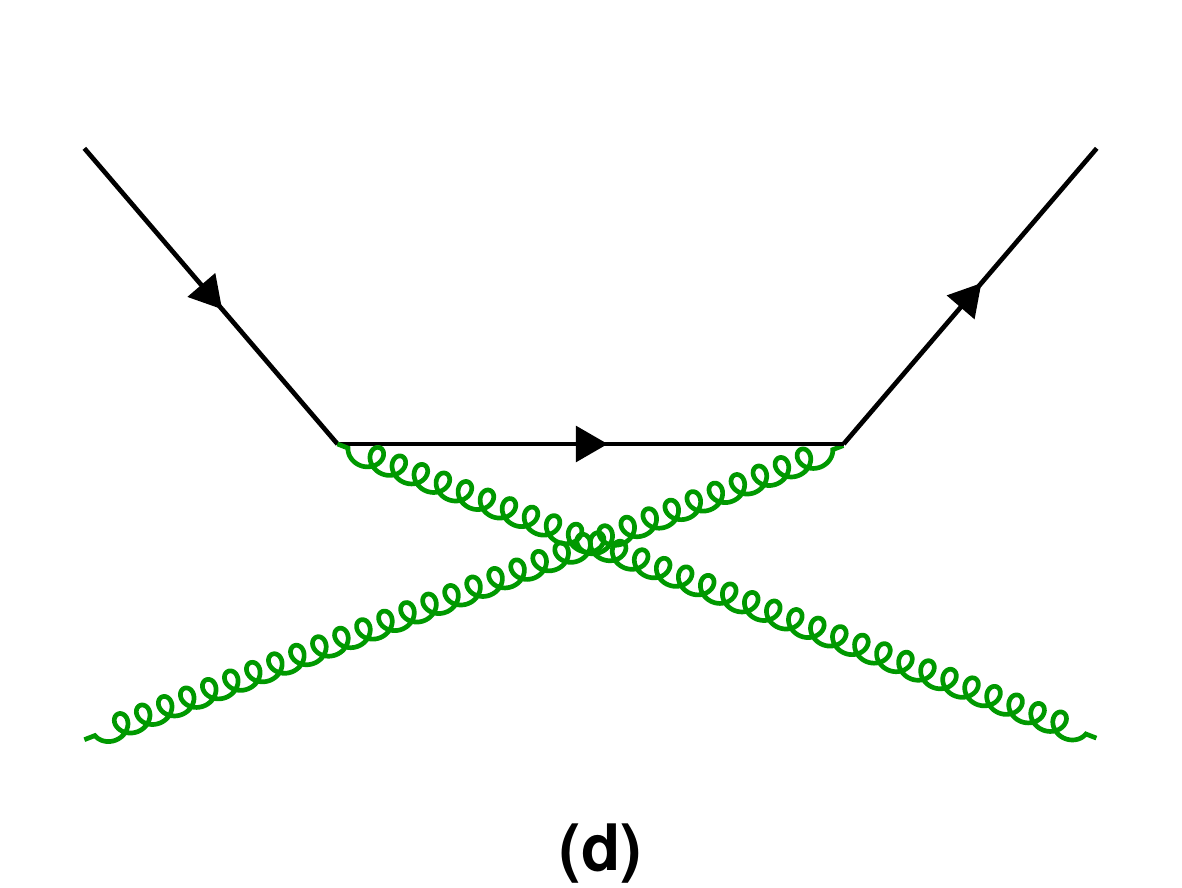}
	\caption{Tree-level Feynman diagrams for the scattering processes $Qq\rightarrow Qq$ (panel-a) and $Qg\rightarrow Qg$ (panel-b, c, d) in vacuum.}
	\label{fig:TwoBody_Diag}
\end{figure}

The elastic scattering processes between heavy quark and the quark-gluon plasma
constituents can be regarded as
\begin{equation}\label{eq:Process}
	\begin{aligned}
		Q~(p_{1})+i~(p_{2})\rightarrow Q~(p_{3})+i~(p_{4}),
	\end{aligned}
\end{equation}
where, $p_{1}=(E_{1},\vec{p}_{1})$ and $p_{2}$ denotes the four-momentum of
heavy quark ($Q$) and the medium partons ($i=q,g$), respectively.
$p_{3}$ and $p_{4}$ are the ones after scattering.
The corresponding tree-level Feynman diagrams for these processes are shown in Fig.~\ref{fig:TwoBody_Diag}.
The quark-quark scattering only has a $t$-channel momentum exchange,
as displayed in the panel-a of Fig.~\ref{fig:TwoBody_Diag},
while the quark-gluon scattering contributes three diagrams corresponding to
$t$-, $s$- and $u$-channels, as presented in the panel-b, c, d in Fig.~\ref{fig:TwoBody_Diag}.
For each diagram in Fig.~\ref{fig:TwoBody_Diag},
the corresponding matrix elements at leading order in $g$
can be found in Appendix-\ref{appendix:Cal_ELoss},
and the four-momentum transfer is
$p^{\mu}_{1}-p^{\mu}_{3} = (\omega,~\vec{q}^{\;}) = (\omega,~\vec{q}_{T},~q_{L})$.
The related Mandelstam invariants can be expressed as
\begin{equation}
	\begin{aligned}\label{eq:MandVar}
		&t \equiv (p_{1}-p_{3})^{2} = \omega^{2} - q^{2} \\
		&s \equiv (p_{1}+p_{2})^{2} \\		
		&u \equiv (p_{1}-p_{4})^{2},
	\end{aligned}
\end{equation}
where, the three-momentum transfer $q\equiv |\vec{q}\;|$.

Since gluons are massless,
the characteristic $t$-channel gluon propagator diverges for
small momentum transfers $t\rightarrow0$, see Eq.~\ref{eq:App_MatrixHQq} and \ref{eq:App_MatrixHQg_t},
which causes a diverging cross-section,
\begin{equation}\label{eq:InfraredDivergence}
	\begin{aligned}
		\frac{d\sigma}{dt} \propto \overline{|\mathcal{M}^{2}|} \propto \frac{1}{t^{2}}.
	\end{aligned}
\end{equation}
The divergence is usually regulated by a cutoff scale for the momentum
phase space, which is encoded in an additional factor~\cite{PhysRevC.82.024906, PhysRevC.91.054908, PhysRevC.94.014909}
\begin{equation}\label{eq:LBTDivergence}
	\begin{aligned}
		\theta(s\geqslant 2m^{2}_{D})\theta(-s+m^{2}_{D}\leqslant t \leqslant -m^{2}_{D}),
	\end{aligned}
\end{equation}
or by a mass $\mu_{D}(T)$ to include the effects of Debye screening~\cite{PhysRevD.79.065039}, 
\begin{equation}\label{eq:ScreenMassDivergence}
	\begin{aligned}
		\frac{1}{t} \rightarrow \frac{1}{t-\mu^{2}_{D}(T)}.
	\end{aligned}
\end{equation}
Note that $\mu_{D}(T)$ is assumed to behave $\mu_{D}(T)\propto T$ in Ref.~\cite{Benjamin88},
and to follow $\mu_{D}^{2}(T)=\lambda m^{2}_{D}(T)$ in the other literatures,
where, $m_{D}\propto T$ is the Debye mass for gluons with a fixed coupling constant, see Eq.~\ref{eq:DebMas} below.
It is argued that, in Ref.~\cite{PhysRevD.26.1394}, $\lambda=1/3$ when taking the thermal gluon mass as the regulator,
while in Ref.~\cite{PBGPRC08}, $\lambda \approx 0.2$ is adjusted requiring that
a pQCD Born calculation with this gluon propagator gives the same
energy loss as the hard-thermal-loop approach.
However, these infrared regulators are not very well determined on first principles.

\begin{figure}[!htbp]
\centering
\includegraphics[width=.27\textwidth]{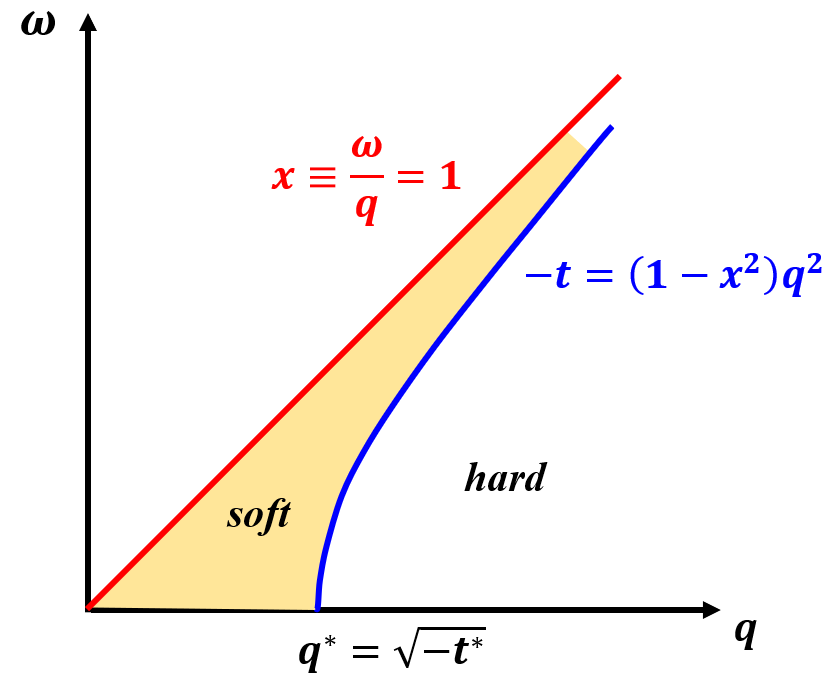}
\caption{Illustration of the phase space decomposed for the soft and hard interactions.}
\label{fig:OmegaQ_Frame}
\end{figure}

Alternatively, the divergence in Eq.~\ref{eq:InfraredDivergence}
can be cured by taking into account the contributions from the long-wavelength gluons,
which correspond to small momentum transfer $\sqrt{-t}\sim gT$, i.e. soft scattering, in a thermal perturbation theory.
The soft gluon exchange in $t$-channels (see panel-a and panel-b in Fig.~\ref{fig:TwoBody_Diag})
features long-range interactions, and they are therefore screened by the medium partons.
Formally, the associated gluon propagator must be screened
with its self-energy~\cite{Braaten91PRL, JeanPR02}.
Concerning the contributions from the large momentum transfer $\sqrt{-t}\gtrsim T$, i.e. hard scattering,
where the Born approximation is valid, and it is straightforward to perform a pQCD calculation in this regime.
This is the soft-hard factorized
approach~\cite{PhysRevD.44.1298, PhysRevD.44.R2625, Romatschke:2003vc, Romatschke:2004au, PhysRevC.74.064907, PhysRevD.77.014015, PhysRevD.77.114017, POWLANGEPJC11},
which allows to decompose the soft HQ-medium interactions with $-t<-t^{\ast}$, from the hard ones with $-t>-t^{\ast}$,
as illustrated in Fig.~\ref{fig:OmegaQ_Frame}.
The intermediate scale $t^{\ast}$ is formally chosen as
\begin{equation}
	\begin{aligned}\label{eq:CutOff_IO}
		&m_{D}^{2}\ll -t^{\ast}\ll T^{2},
	\end{aligned}
\end{equation}
implying that the coupling is very small $\alpha_{s}\ll1$,
namely the weak-coupling or high-temperature limit~\cite{Braaten91PRL, Romatschke:2004au, PBGPRC08}.
This relation, strictly speaking,
should be guaranteed if applied to realistic situations
where the temperature is not high enough
and the coupling is not terribly small.
However, in our calculations, we simply explore the weak-coupling limit
and see what they give in practice.
In the previous work, 
we calculated the transport coefficients in this limit and especially studied how they change within the temperature region accessed by the RHIC and LHC experiments.
By comparing the results based on a set of the intermediate cutoff $t^{\ast}$,
it is found that, the momentum diffusion coefficients
behave a mild sensitivity~\cite{Li_2021},
supporting the validity of the soft-hard factorization approach at RHIC and LHC energies.
We will check further this conclusion via the energy loss in this work.
We note that the divergence in Eq.~\ref{eq:InfraredDivergence} is
usually regulated by the sharp cutoff on the momentum transfer, such as $t^{\ast}$,
which can also be cured by a dimensional regularization.
See Ref.~\cite{PhysRevD.103.116002} for details.

\subsection{The interaction rate for two-body scattering}\label{subsec:gamma_HTLpQCD}
In this sub-section, we need to evaluate the scattering rate from the proceeding calculations.
With the soft-hard factorization model,
the scattering rate arises from three kinematic regimes:
\begin{enumerate}
	\item[(1)] The scattering of heavy quarks and medium partons in $t$-channel with small momentum transfer, $-t<-t^{\ast}$.
	This is the ``$soft$'' region.
	\item[(2)] The scattering in $t$-channel with large momentum transfer, $-t>-t^{\ast}$. This is the ``$hard$''region.
	\item[(3)] The scattering of heavy quarks and gluons in $s$- and $u$-channels with both small and large momentum transfer.
	This is the ``$s+u$'' region, whose contribution is limited since the relevant interaction rate is
	much smaller when compared with the one from $t$-channel.
\end{enumerate}
The total scattering rate is the sum of these three kinematic regions
\begin{equation}
	\begin{aligned}\label{eq:Gamma_Full}
		\Gamma = \Gamma^{soft}_{(t)} + \Gamma^{hard}_{(t)} + \Gamma_{(s+u)},
	\end{aligned}
\end{equation}
in which the expressions of each of these regions will be summarized below.
Here we just show the final results obtained in our previous work, and the details are relegated to Ref.~\cite{Li_2021} and the references therein.

In soft collisions, the scattering rate for two-body scattering is expressed as
\begin{equation}
\begin{aligned}\label{eq:Gamma_Soft}
{\Gamma}^{soft}_{(t)}(E_{1},T) = &C_{F}g^{2} \int_{q} \int d\omega \; \bar{n}_{B}(\omega) \delta(\omega-\vec{v}_{1}\cdot\vec{q}\;) \\
&\biggr\{ \rho_{L}(\omega,q) + \vec{v}_{1}^{\;2} \bigr[ 1-(\hat{v}_{1}\cdot\hat{q})^{2} \bigr]\rho_{T}(\omega,q) \biggr\}
\end{aligned}
\end{equation}
where, $C_{F}=4/3$ is the quark Casimir factor;
$g$ is the strong coupling constant~\cite{TwoLoopGPRD05}
\begin{equation}
\begin{aligned}\label{eq:TwoLoopG}
&g^{-2}(\mu)=2\beta_{0} ln(\frac{\mu}{\Lambda_{\rm QCD}}) + \frac{\beta_{1}}{\beta_{0}} ln \biggr[ 2ln(\frac{\mu}{\Lambda_{\rm QCD}}) \biggr]
\end{aligned}
\end{equation}
where, the factors $\beta_{0}=(11-\frac{2}{3}N_{f})/(16\pi^{2})$, $\beta_{1}=(102-\frac{38}{3}N_{f})/(16\pi^{2})^{2}$
and the scales $\mu=\pi T$, $\Lambda_{\rm QCD}=261~{\rm MeV}$;
$N_{f}$ is the number of active flavors in the quark-gluon plasma;
$\vec{v}_{1}=\vec{p}_{1}/E_{1}$ is the heavy-quark velocity;
$n_{B/F}(E)=(e^{E/T}\mp1)^{-1}$ denotes the thermal distributions for Bosons/Fermions
and $\bar{n}_{B/F}\equiv 1 \pm n_{B/F}$ accounts for the Bose-enhancement or Pauli-blocking effect.
Note that the short notation $\int_{q}\equiv \int\frac{d^{3}\vec{q}}{(2\pi)^{3}}$
is adopted for momentum phase space integrals.
The transverse and longitudinal spectral functions $\rho_{T/L}$ in Eq.~\ref{eq:Gamma_Soft} reads
\begin{equation}
\begin{aligned}\label{eq:GammaRhoT_Soft1}
\rho_{T}(\omega,q) = &\frac{\pi \omega m_{D}^{2}}{2q^{3}} (q^{2}-\omega^{2}) \biggr\{ \biggr[ q^{2}-\omega^{2} + \frac{\omega^{2}m_{D}^{2}}{2q^{2}} \\
&\bigr(1+\frac{q^{2}-\omega^{2}}{2\omega q} ln\frac{q+\omega}{q-\omega}\bigr) \biggr]^{2} \\
&+ \biggr[ \frac{\pi \omega m_{D}^{2}}{4q^{3}} (q^{2}-\omega^{2}) \biggr]^{2} \biggr\}^{-1},
\end{aligned}
\end{equation}
\begin{equation}
\begin{aligned}\label{eq:GammaRhoL_Soft1}
\rho_{L}(\omega,q) = &\frac{\pi \omega m_{D}^{2}}{q}
\biggr\{ \biggr[ q^{2}+m_{D}^{2}\bigr(1-\frac{\omega}{2q} ln\frac{q+\omega}{q-\omega}\bigr) \biggr]^{2} \\
&+ \biggr( \frac{\pi \omega m_{D}^{2}}{2q} \biggr)^{2} \biggr\}^{-1},
\end{aligned}
\end{equation}
in which $m_{D}^{2}$ is the Debye screening mass squared for gluon
\begin{equation}
\begin{aligned}\label{eq:DebMas}
m_{D}^{2}=\bigr( \frac{N_{c}}{3} + \frac{N_{f}}{6} \bigr)g^{2} T^{2},
\end{aligned}
\end{equation}
where, $N_{c}=3$ is the color factor and $N_{f}=3$ is the quark flavors.

In hard collisions of heavy quark ($Q$) and medium parton ($i=q,g$), the corresponding scattering rate in $t$-channel reads
\begin{equation}
\begin{aligned}\label{eq:Gamma_Hard}
{\Gamma}^{hard}_{Qi(t)}(E_{1},T) = &\frac{1}{2E_{1}} \int_{p_{2}} \frac{n(E_{2})}{2E_{2}}
\int_{p_{3}} \frac{1}{2E_{3}} \int_{p_{4}} \frac{\bar{n}(E_{4})}{2E_{4}} \\
&\overline{|\mathcal{M}^{2}|}_{Qi(t)} (2\pi)^{4} \delta^{(4)}(p_{1}+p_{2}-p_{3}-p_{4}),
\end{aligned}
\end{equation}
which is obtained by neglecting the thermal effects on the heavy quark after scattering.
Similarly, the hard contributions from the scattering of heavy quarks and gluons in $s$- and $u$-channels
can be obtained by modifying Eq.~\ref{eq:Gamma_Hard} as
\begin{equation}
	\begin{aligned}\label{eq:Gamma_su}
		{\Gamma}^{hard}_{Qg(s+u)}(E_{1},T) = &\frac{1}{2E_{1}} \int_{p_{2}} \frac{n(E_{2})}{2E_{2}}
		\int_{p_{3}} \frac{1}{2E_{3}} \int_{p_{4}} \frac{\bar{n}(E_{4})}{2E_{4}} \\
		&\overline{|\mathcal{M}^{2}|}_{Qg(s+u)} (2\pi)^{4} \delta^{(4)}(p_{1}+p_{2}-p_{3}-p_{4}).
	\end{aligned}
\end{equation}

\subsection{The collisional energy loss for two-body scattering}\label{subsec:eloss_HTLpQCD}
The heavy quark energy loss per distance traveled is written as
\begin{equation}\label{eq:ELoss_Def}
	\begin{aligned}
		-\frac{dE}{dz}=\int d^{3}\vec{q} \; \frac{d\Gamma}{d^{3}\vec{q}} \; \frac{\omega}{v_{1}}
	\end{aligned}
\end{equation}
where, $v_{1}$ is the velocity of the heavy quark,
$d\Gamma/d^{3}\vec{q}$ is the differential scattering rate with respect to the three-momentum transfer $\vec{q}$
and $\omega$ is the energy transfer.
The total energy loss in the soft-hard factorized approach is given by
inserting Eq.~\ref{eq:Gamma_Full} into Eq.~\ref{eq:ELoss_Def}, yielding
\begin{equation}
\begin{aligned}\label{eq:dEdz_Full}
-\frac{dE}{dz} = \biggr[-\frac{dE}{dz}\biggr]^{soft}_{(t)} + \biggr[-\frac{dE}{dz}\biggr]^{hard}_{(t)} + \biggr[-\frac{dE}{dz}\biggr]_{(s+u)},
\end{aligned}
\end{equation}
in which the contributions from $t$-channel are shown as the first ($soft$) and second ($hard$) terms on the right hand side,
together with the contributions from $s$- and $u$-channels are expressed as the third term.

With Eq.~\ref{eq:Gamma_Soft} we can get the energy loss in soft collisions,
\begin{equation}
\begin{aligned}\label{eq:dEdz_Soft}
\biggr[-\frac{dE}{dz}\biggr]^{soft}_{(t)} = &\frac{C_{F}g^{2}}{8\pi^{2}v^{2}_{1}} \int^{0}_{t^{*}} dt \; (-t) \int_{0}^{v_{1}}dx \frac{x}{(1-x^{2})^{2}} \\
&\bigr[ \rho_{L}(t,x) + (v_{1}^{2}-x^{2})\rho_{T}(t,x) \bigr],
\end{aligned}
\end{equation}
in which the integration variables ($t=\omega^2-q^{2},x=\omega/q$) are changed
comparing with the ones ($q,\omega$) as shown in Eq.~\ref{eq:Gamma_Soft}.
See Fig.~\ref{fig:OmegaQ_Frame} for details.
Thus, the transverse and longitudinal spectral functions (Eq.~\ref{eq:GammaRhoT_Soft1} and \ref{eq:GammaRhoL_Soft1})
can be rewritten as~\cite{Li_2021}
\begin{equation}
\begin{aligned}\label{eq:GammaRhoT_Soft2}
\rho_{T}(t,x) = &\frac{\pi m_{D}^{2}}{2} x(1-x^{2}) \biggr\{ \biggr[-t+\frac{m_{D}^{2}}{2} x^{2} \bigr(1 + \frac{1-x^{2}}{2x} \\
&ln\frac{1+x}{1-x} \bigr) \biggr]^{2} + \biggr[ \frac{\pi m_{D}^{2}}{4}x(1-x^{2}) \biggr]^{2} \biggr\}^{-1},
\end{aligned}
\end{equation}
\begin{equation}
\begin{aligned}\label{eq:GammaRhoL_Soft2}
\rho_{L}(t,x) = &\pi m_{D}^{2} x \biggr\{ \biggr[ \frac{-t}{1-x^{2}}+m_{D}^{2} \bigr(1-\frac{x}{2} \\
&ln\frac{1+x}{1-x} \bigr) \biggr]^{2} + \biggr( \frac{\pi m_{D}^{2}}{2} x \biggr)^{2} \biggr\}^{-1}.
\end{aligned}
\end{equation}

In analogy with the soft part,
the energy loss in hard collisions can be obtained with Eq.~\ref{eq:Gamma_Hard},
\begin{equation}
\begin{aligned}\label{eq:dEdz_Hard}
\biggr[-\frac{dE}{dz}\biggr]^{hard}_{(t)} =& \sum_{i=q,g}\biggr[-\frac{dE}{dz}\biggr]^{hard}_{Qi(t)} \\
=& \frac{1}{256\pi^{3}\vec{p}_{1}^{\;2}} \sum_{i=q,g}\int_{|\vec{p}_{2}|_{min}}^{\infty}d|\vec{p}_{2}| ~E_{2} n(E_{2}) \\
&\int_{-1}^{cos\psi|_{max}} d(cos\psi) \int_{t_{min}}^{t^{*}}dt \frac{b}{a^{3}} \; \overline{|\mathcal{M}^{2}|}_{Qi(t)}.
\end{aligned}
\end{equation}
The boundaries of the integrals ($|\vec{p}_{2}|_{min}$, $cos\psi|_{max}$ and $t_{min}$),
the parameters ($a$ and $b$) and the matrix elements in vacuum ($\overline{|\mathcal{M}^{2}|}_{Qi}$) in Eq.~\ref{eq:dEdz_Hard},
are expressed and summarized in Appendix-\ref{appendix:Cal_ELoss}.
More detailed aspects of the calculations are also shown in this appendix.

Due to the finite heavy quark mass ($m_{1}$ in a few times $\rm GeV$),
the contributions from $s$- and $u$-channels, $[dE/dz]_{(s+u)}$ in Eq.~\ref{eq:dEdz_Full},
are not divergent for small momentum transfers.
Thus, there is no need to introduce the intermediate cutoff $t^{\ast}$
and the $|\vec{p}_{2}|$ integration in Eq.~\ref{eq:dEdz_Hard} can be continued down to zero,
as well as the $cos\psi$ and $t$ integrations can be continued up to unity and zero, respectively,
resulting in
\begin{equation}
\begin{aligned}\label{eq:dEdz_su}
\biggr[-\frac{dE}{dz}\biggr]_{(s+u)} =& \frac{1}{256\pi^{3}\vec{p}_{1}^{\;2}} \int_{0}^{\infty}d|\vec{p}_{2}| ~E_{2} n(E_{2}) \\
&\int_{-1}^{1} d(cos\psi) \int_{t_{min}}^{0}dt \frac{b}{a^{3}} \; \overline{|\mathcal{M}^{2}|}_{Qg(s+u)}.
\end{aligned}
\end{equation}

\subsection{The energy loss in the high-energy approximation}\label{subsec:approximation}
The integrals in Eqs.~\ref{eq:dEdz_Soft}, \ref{eq:dEdz_Hard} and \ref{eq:dEdz_su} are difficult
to evaluate analytically for arbitrary heavy quark energy $E_{1}$.
The physical interpretation of the results is challenging in general.
However, in the high-energy approximation (HEA, i.e. $E_{1}\rightarrow\infty$),
the relevant results are surprisingly simple,
and they are certainly useful to discuss further the in-medium energy loss mechanisms.
Here, we simply summarize the final results,
and the detailed aspects are relegated to Appendix-\ref{appendix:Cal_HighEnergyLimit}.

Concerning the soft contribution ($-t<-t^{\ast}$)
in the limit $E_{1}\rightarrow \infty$ we are interested in,
the full result (Eq.~\ref{eq:dEdz_Soft}) can be simplified as (Eq.~\ref{eq:App_dEdx_Soft_HEA})
\begin{equation}
        \begin{aligned}\label{eq:dEdx_Soft_HEA}
                &\biggr[-\frac{dE}{dz}\biggr]^{soft-HEA}_{(t)} = \frac{C_{F}}{16\pi} \biggr( \frac{N_{c}}{3}+\frac{N_{f}}{6} \biggr) g^{4} T^{2} ln\frac{-2t^{\ast}}{m_{D}^{2}}.
        \end{aligned}
\end{equation}
It appears that (1) $dE/dz$ depends logarithmically on the intermediate cutoff $t^{\ast}$;
(2) its temperature-dependency behaves $dE/dz\propto T^{2}ln(T+const.)$ at fixed coupling;
(3) its energy-dependency vanishes.

For hard collisions ($-t>-t^{\ast}$) with very hard momentum exchange,
the kinematics are constrained by
$-t \approx s \approx \tilde{s} \sim \mathcal{O}(E_{1}T)$
or $-\tilde{u} \ll \tilde{s} \sim \mathcal{O}(E_{1}T)$
with the abbreviations $\tilde{s}\equiv s-m^{2}_{1}$ and $\tilde{u}\equiv u-m^{2}_{1}$.
Accordingly, the energy loss for $Qq$ and $Qg$ scatterings in
different channels (Eqs.~\ref{eq:dEdz_Hard} and \ref{eq:dEdz_su}) can be simplified as
(Eqs.~\ref{eq:App_dEdx_Hard_HEA_Qq}, \ref{eq:App_dEdx_Hard_HEA_Qg_t} and \ref{eq:App_dEdx_Hard_HEA_Qg_su})
\begin{equation}
        \begin{aligned}\label{eq:dEdx_Hard_HEA_Qq}
                &\biggr[-\frac{dE}{dz}\biggr]^{hard-HEA}_{Qq(t)} = \frac{N_{f}N_{c}}{216\pi} g^{4}T^{2} \biggr( ln \frac{8E_{1}T}{-t^{\ast}}-\frac{3}{4}+c \biggr),
        \end{aligned}
\end{equation}
\begin{equation}
        \begin{aligned}\label{eq:dEdx_Hard_HEA_Qg_t}
                &\biggr[-\frac{dE}{dz}\biggr]^{hard-HEA}_{Qg(t)} = \frac{N_{c}^{2}-1}{96\pi} g^{4} T^{2} \biggr( ln \frac{4E_{1}T}{-t^{\ast}}-\frac{3}{4}+c \biggr),
        \end{aligned}
\end{equation}
\begin{equation}
        \begin{aligned}\label{eq:dEdx_Hard_HEA_Qg_su}
                &\biggr[-\frac{dE}{dz}\biggr]^{hard-HEA}_{Qg(s+u)} = \frac{N_{c}^{2}-1}{432\pi} g^{4} T^{2} \biggr( ln \frac{4E_{1}T}{m^{2}_{1}}-\frac{5}{6}+c \biggr),
        \end{aligned}
\end{equation}
where, the strong coupling constant $g^{2}=4\pi\alpha_{s}$ and the constant factor $c\approx -1.14718$.
It is found that (1) the logarithmic terms $ln[E_{1}T/(-t^{\ast})]$ and $ln(E_{1}T/m^{2}_{1})$ arising from the integral
$\int dt/t$ ($t$-channel) and $\int d\tilde{u}/\tilde{u}$ ($u$-channel), respectively,
in the region $-t \approx \tilde{s} \gg m^{2}_{1}$ when $s \approx \tilde{s}\sim \mathcal{O}(E_{1}T)$;
(2) $dE/dz$ depends on the intermediate scale $t^{\ast}$ for the contribution from $t$-channel,
while it is not for the $s$- and $u$-channels;
(3) its temperature-dependency behaves $dE/dz\propto T^{2}(lnT+const.)$ for a given energy;
(4) its energy-dependency behaves $dE/dz\propto lnE_{1}$ for a given temperature at fixed coupling;
(5) $dE/dz$ for the $Qg$ scatterings in $t$-channel is much larger that in $s$- and $u$-channels.
The temperature and energy dependencies of $dE/dz$ are
similar to the results for the scattering of a light hard parton
off a light soft parton~\cite{PhysRevC.71.034901, PhysRevLett.100.072301}.

Summing all these contributions up,
we obtain the total energy loss loss of heavy quark from scattering off
quarks and gluons in the high-energy approximation
\begin{equation}
        \begin{aligned}\label{eq:dEdx_Soft_Hard_HEA}
                \biggr[-\frac{dE}{dz}\biggr]^{HEA}_{Qq+Qg} =& \frac{4}{3}\pi\alpha_{s}^{2}T^{2} \biggr[ \bigr( 1 + \frac{N_{f}}{6} \bigr) ln\frac{E_{1}T}{m_{D}^{2}} \\
		& + \frac{2}{9}ln\frac{E_{1}T}{ m^{2}_{1}} + d(N_{f}) \biggr],
        \end{aligned}
\end{equation}
where, $N_{c}=3$ and $d(N_{f})\approx 0.145901N_{f}+0.050213$.
It is realized that the dependence on the arbitrary scale $t^{\ast}$ cancels,
which is similar to the QED case~\cite{PhysRevD.44.1298, PhysRevD.77.014015}.
Same results can be found in Ref.~\cite{PhysRevD.77.114017}.

\subsection{The energy loss calculated with the Einstein's relationship}\label{subsec:eloss_ER}
During the traversing through the QCD medium,
the heavy quark dynamics is usually described by the Boltzmann model.
For the Boltzmann approach,
it is argued~\cite{PhysRevC.71.064904} that the interactions between heavy quark and medium partons
can be conveniently encoded into the drag ($\eta_{D}$) and momentum diffusion coefficients ($\kappa_{T}$ and $\kappa_{L}$),
which describe, respectively, the average energy loss and the momentum fluctuations
in the direction that is parallel and perpendicular to the propagation.
All the transport coefficients can be calculated independently.
Assuming a small momentum transfer in interactions,
Boltzmann is reduced to the Fokker-Plank dynamics,
which can be realized stochastically by a Langevin approach~\cite{Benjamin88, HQQGPRapp10, Zhao:2020jqu, He:2022ywp}.

In the framework of the Fokker-Planck and Langevin transport~\cite{Akamatsu09,HFModelHee13,CaoPRC15,CTGUHybrid1,CTGUHybrid2,CTGUHybrid3},
the drag and momentum diffusion coefficients are related to each other
via the Einstien fluctuation-dissipation relation.
Consequently, the heavy quarks allow to reach the thermodynamic equilibrium
and follow a Maxwell-Boltzman distribution in the infinite time limit~\cite{CaoPRC13}.
The drag coefficient can be characterized by the energy loss per length,
$-\eta_{D} |\vec{p}^{\;}| = d|\vec{p}^{\;}|/dt=dE/(vdt)=dE/dz$, yielding
\begin{equation}
	\begin{aligned}\label{eq:ELoss_EtaD_pQCDHTL}
		-\frac{dE}{dz} = \eta_{D}|\vec{p}_{1}|.
	\end{aligned}
\end{equation}
The drag coefficient $\eta_{D}$ in Eq.~\ref{eq:ELoss_EtaD_pQCDHTL} can be given via the Einstein's relationship~\cite{CTGUHybrid4}
\begin{equation}
        \begin{aligned}\label{eq:EtaD_DissFluc}
                \eta_{D} =& \eta_{D}(\kappa_{T}, \kappa_{L}) \\
                =& \frac{\kappa_{L}}{2TE_{1}} + (\xi-1)\frac{\partial \kappa_{L}}{\partial \vec{p}_{1}^{\;2}} + \frac{1}{\vec{p}_{1}^{\;2}}  \biggr[ \xi(\sqrt{\kappa_{T}}+\sqrt{\kappa_{L}})^{2} \\
                &- (3\xi-1)\kappa_{T} - (\xi+1)\kappa_{L} \biggr],
        \end{aligned}
\end{equation}
in which the parameter $\xi=0,0.5,1$, corresponding to the pre-point Ito, the mid-point Stratonovic
and the post-point discretization schemes, respectively~\cite{HQQGPRapp10}.
The post-point Ito scheme ($\xi=1$) is adopted in this work.

We can see that now all the associated components, such as the drag coefficient and the energy loss,
are quantified by the transverse and longitudinal momentum diffusion coefficients~\cite{Li_2021},
\begin{equation}
        \begin{aligned}\label{eq:KappaT}
                \kappa_{T} &=\frac{1}{2}  \int d^{3}\vec{q} \; \frac{d\Gamma}{d^{3}\vec{q}} \biggr[ \omega^{2} - t - \frac{(2\omega E_{1}-t)^{2}}{4\vec{p}_{1}^{\;2}} \biggr]
        \end{aligned}
\end{equation}
\begin{equation}
        \begin{aligned}\label{eq:KappaL}
                \kappa_{L} &= \frac{1}{4\vec{p}_{1}^{\;2}} \int d^{3}\vec{q} \frac{d\Gamma}{d^{3}\vec{q}} \; (2\omega E_{1}-t)^{2},
        \end{aligned}
\end{equation}
which can be formulated with the soft-hard factorization approach,
\begin{equation}
	\begin{aligned}\label{eq:KappaT_pQCDHTL}
		\kappa_{T} &= \bigr(\kappa_{T}\bigr)^{soft}_{(t)} + \bigr(\kappa_{T}\bigr)^{hard}_{(t)} +\bigr(\kappa_{T}\bigr)_{(s+u)},
	\end{aligned}
\end{equation}
\begin{equation}
	\begin{aligned}\label{eq:KappaL_pQCDHTL}
		\kappa_{L} &= \bigr(\kappa_{L}\bigr)^{soft}_{(t)} + \bigr(\kappa_{L}\bigr)^{hard}_{(t)} + \bigr(\kappa_{L}\bigr)_{(s+u)}.
	\end{aligned}
\end{equation}
The soft components read~\cite{Li_2021}
\begin{equation}
	\begin{aligned}\label{eq:KappaT_Soft}
		\bigr(\kappa_{T}\bigr)^{soft}_{(t)} =& \frac{C_{F}g^{2}}{16\pi^{2}v_{1}^{3}} \int^{0}_{t^{\ast}} dt
		\int_{0}^{v_{1}}dx ~\mathcal{A} \cdot (v_{1}^{2}-x^{2}),
	\end{aligned}
\end{equation}
\begin{equation}
	\begin{aligned}\label{eq:KappaL_Soft}
		\bigr(\kappa_{L}\bigr)^{soft}_{(t)}=&\frac{C_{F}g^{2}}{8\pi^{2}v_{1}^{3}} \int^{0}_{t^{\ast}} dt
		\int_{0}^{v_{1}}dx ~\mathcal{A} \cdot x^{2},
	\end{aligned}
\end{equation}
where,
\begin{equation}
	\begin{aligned}
		\mathcal{A} \equiv & (-t)^{\frac{3}{2}} \biggr[ \rho_{L}(t,x) + (v_{1}^{2}-x^{2})\rho_{T}(t,x) \biggr] \\
		& \times coth\biggr(\frac{x}{2T}\sqrt{\frac{-t}{1-x^{2}}} \biggr) \biggr/ (1-x^{2})^{\frac{5}{2}}.
	\end{aligned}
\end{equation}
The hard contributions from $t$-channel take the form~\cite{Li_2021}
\begin{equation}
	\begin{aligned}\label{eq:KappaT_Hard_t}
		\bigr(\kappa_{T}\bigr)^{hard}_{(t)} =& \sum_{i=q,g} \bigr(\kappa_{T}\bigr)^{hard}_{Qi(t)} \\
		=& \frac{1}{256\pi^{3}|\vec{p}_{1}|E_{1}} \sum_{i=q,g} \int_{|\vec{p}_{2}|_{min}}^{\infty}d|\vec{p}_{2}| E_{2} n_{2}(E_{2}) \\
		&\int_{-1}^{cos\psi|_{max}} d(cos\psi) \int_{t_{min}}^{t^{\ast}}dt ~\mathcal{B} \cdot \overline{|\mathcal{M}^{2}|}_{Qi(t)}
	\end{aligned}
\end{equation}
\begin{equation}
	\begin{aligned}\label{eq:KappaL_Hard_t}
		\bigr(\kappa_{L}\bigr)^{hard}_{(t)} =& \sum_{i=q,g} \bigr(\kappa_{L}\bigr)^{hard}_{Qi(t)} \\
		=& \frac{1}{256\pi^{3}|\vec{p}_{1}|^{3}E_{1}} \sum_{i=q,g} \int_{|\vec{p}_{2}|_{min}}^{\infty}d|\vec{p}_{2}| E_{2} n_{2}(E_{2}) \\
		&\int_{-1}^{cos\psi|_{max}} d(cos\psi) \int_{t_{min}}^{t^{\ast}} dt ~\mathcal{C} \cdot \overline{|\mathcal{M}^{2}|}_{Qi(t)},
	\end{aligned}
\end{equation}
where,
\begin{equation}
	\begin{aligned}
		\mathcal{B} &\equiv \frac{1}{a} \biggr[ -\frac{m_{1}^{2}(D+2b^{2})}{8\vec{p}_{1}^{\;2}a^{4}} + \frac{E_{1}tb}{2\vec{p}_{1}^{\;2}a^{2}} - t(1+\frac{t}{4\vec{p}_{1}^{\;2}}) \biggr], \\
		\mathcal{C} &\equiv \frac{1}{a} \biggr[ \frac{E_{1}^{2}(D+2b^{2})}{4a^{4}} -\frac{E_{1}tb}{a^{2}} + \frac{t^{2}}{2} \biggr].
	\end{aligned}
\end{equation}
Note that the boundaries of the integrals ($|\vec{p}_{2}|_{min}$, $cos\psi|_{max}$ and $t_{min}$)
and the parameters ($a$, $b$ and $D$) are given in Appendix-\ref{appendix:Cal_ELoss}.
Similar with Eqs.~\ref{eq:dEdz_Hard} and \ref{eq:dEdz_su},
the hard contributions from $s$- and $u$-channels
can be obtained by modifying the boundaries of the integrals in Eqs.~\ref{eq:KappaT_Hard_t} and \ref{eq:KappaL_Hard_t}, yielding
\begin{equation}
        \begin{aligned}\label{eq:KappaT_su}
                \bigr(\kappa_{T}\bigr)_{(s+u)} =& \frac{1}{256\pi^{3}|\vec{p}_{1}|E_{1}} \int_{0}^{\infty}d|\vec{p}_{2}| E_{2} n_{2}(E_{2}) \\
                &\int_{-1}^{1} d(cos\psi) \int_{t_{min}}^{0}dt ~\mathcal{B} \cdot \overline{|\mathcal{M}^{2}|}_{Qg(s+u)}
        \end{aligned}
\end{equation}
\begin{equation}
        \begin{aligned}\label{eq:KappaL_su}
                \bigr(\kappa_{L}\bigr)_{(s+u)} =& \frac{1}{256\pi^{3}|\vec{p}_{1}|^{3}E_{1}} \int_{0}^{\infty}d|\vec{p}_{2}| E_{2} n_{2}(E_{2}) \\
                &\int_{-1}^{1} d(cos\psi) \int_{t_{min}}^{0} dt ~\mathcal{C} \cdot \overline{|\mathcal{M}^{2}|}_{Qg(s+u)}.
        \end{aligned}
\end{equation}
See Ref.~\cite{Li_2021} for more details.

\section{Results and discussions}\label{sec:result}
\subsection{The collisional energy loss}
\begin{figure*}[!htbp]
\begin{center}
\setlength{\abovecaptionskip}{-0.1mm}
\includegraphics[width=.48\textwidth]{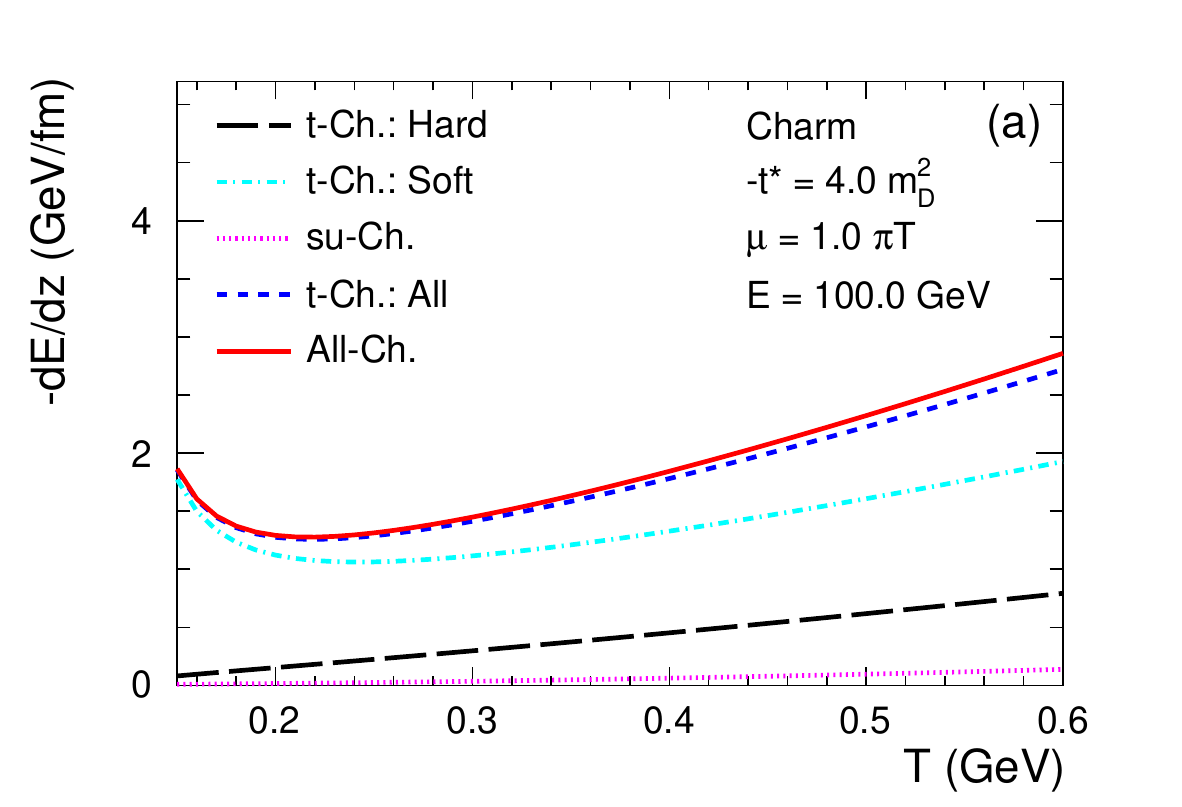}
\includegraphics[width=.48\textwidth]{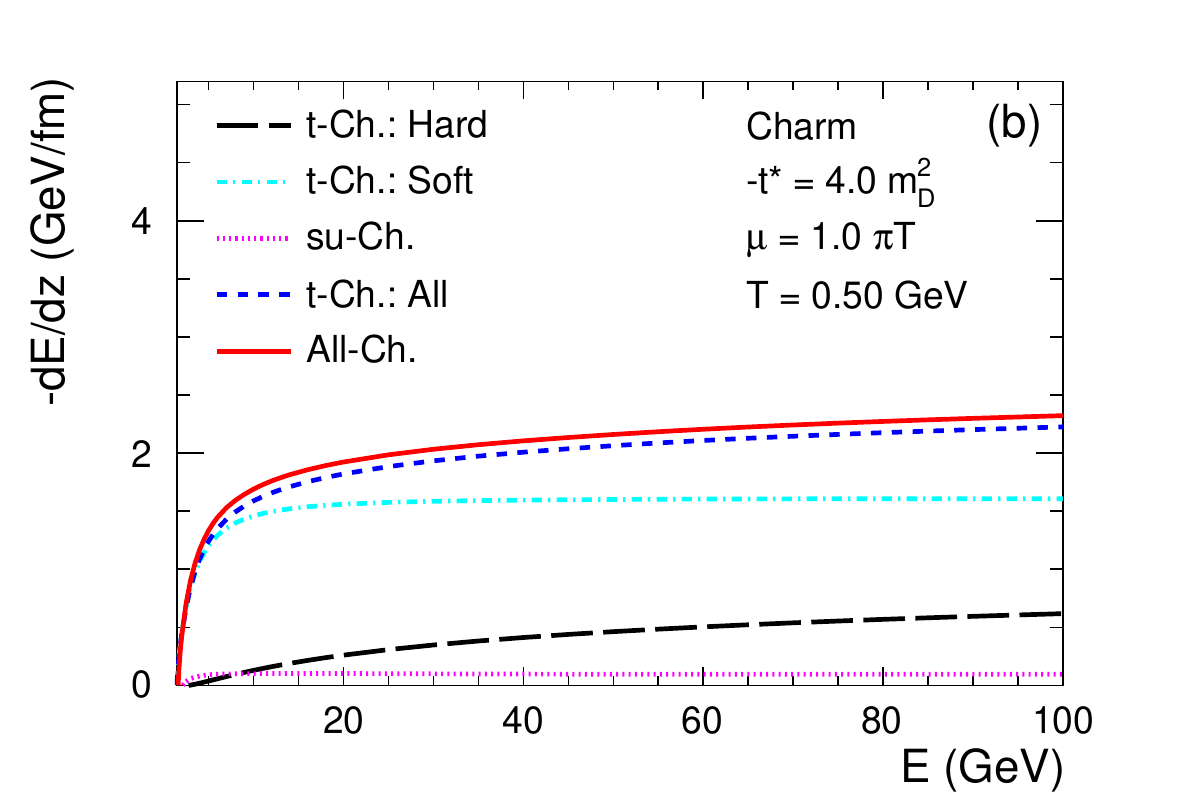}
\caption{(Color online) Left (a): comparison of the energy loss $dE/dz$ as a function of temperature,
for charm quark with $-t^{\ast}=4m_{D}^{2}$ and $\mu=\pi T$ at a given energy $E=100~{\rm GeV}$,
contributed by hard interactions in $t$-channel (long dashed black curve; Eq.~\ref{eq:dEdz_Hard}),
soft interactions in $t$-channel (dot-dashed cyan curve; Eq.~\ref{eq:dEdz_Soft}) and
$s$- and $u$-channels (dotted pink curve; Eq.~\ref{eq:dEdz_su}).
The combined results, i.e. the contributions from the soft and hard interactions in $t$-channels (dashed blue curve)
and from the all components (solid red curve), are shown for comparison.
Right (b): same as panel-a but for $dE/dz$ as a function of heavy quark energy at a given temperature $T=0.5~{\rm GeV}$.}
\label{fig:Charm_dEdz_Full}
\end{center}
\end{figure*}

In Fig.~\ref{fig:Charm_dEdz_Full}(a), the charm quark energy loss obtained with
$-t^{\ast}=4m_{D}^{2}$ and $\mu=\pi T$ at a given initial energy $E=100~{\rm GeV}$,
are presented as a function of QCD medium temperature.
The contributions of the various sources are displayed as curves with different styles.
See the legend for details.
It is found that (1) the soft contribution (dot-dashed cyan curve; Eq.~\ref{eq:dEdz_Soft})
shows a decreasing behavior in the range $T_{c}<T\lesssim0.25~{\rm GeV}$,
following by an increasing trend at higher temperature.
This non-monotonic behavior is mainly induced by the running of the coupling with temperature.
See Fig.~\ref{fig:Charm_dEdz_FullHEA}(a) for details.
When increasing the local temperature,
a heavy quark will suffer more frequent momentum kicks
from its surrounding medium partons, resulting in a stronger interaction strength,
which, in turn, loses more of its initial energy;
(2) the hard contribution from $t$-channel (long dashed black curve; Eq.~\ref{eq:dEdz_Hard})
grows with temperature since the thermal parton distribution function behaves
$n\sim e^{-E/T}$ (Eq.~\ref{eq:App_ThermalDis}) at large energy ($E\gg T$);
(3) the contribution from $s$- and $u$-channels (dotted pink curve; Eq.~\ref{eq:dEdz_su}) exhibits a similar trend.
Comparing with the result from $t$-channel (dashed blue curve),
this contribution is negligible,
which is expected since the relevant interaction rate is much smaller with respect to the one in $t$-channel;
(4) the soft contribution is close to the combined result (solid red curve),
reflecting its dominance in the whole temperature range.
We note that the soft (hard) contribution to the energy loss increases (decreases) with the scale $|t^{\ast}|$.

Figure~\ref{fig:Charm_dEdz_Full}(b) shows the results
for $T=0.5~{\rm GeV}$ as a function of heavy quark energy.
All the components show a monotonously rising energy dependence.
The soft contribution (dot-dashed cyan curve) dominates in the considered energy region.
It has a stronger energy dependence at low energy $E\lesssim 3~{\rm GeV}$,
followed by an almost flat behavior at much higher energy $E\gtrsim 10~{\rm GeV}$.
This may be induced by the fact that,
compared with the diffusion term,
the drag term dominates the scatterings since the initial
momentum spectra of heavy quarks is much harder than that of medium partons.
Thus, the relevant in-medium energy loss can be described by the drag force,
which is proportional to (a positive power of) the heavy quark velocity $v=\sqrt{E^{2}-m^{2}_{Q}}/E$.
The relativistic effect is trivial in the low-energy region
($m_{Q}\leqslant E \lesssim2m_{Q}$), where the velocity and the energy loss change significantly with increasing energy.
However, the ultrarelativistic effect should be considered in the very large energy region ($E\gg m_{Q}$),
where the velocity is very close to unity and consequently,
the energy loss will increase very slowly.
Same conclusions can be drawn for bottom quarks.

\begin{figure*}[!htbp]
\begin{center}
\setlength{\abovecaptionskip}{-0.1mm}
\includegraphics[width=.47\textwidth]{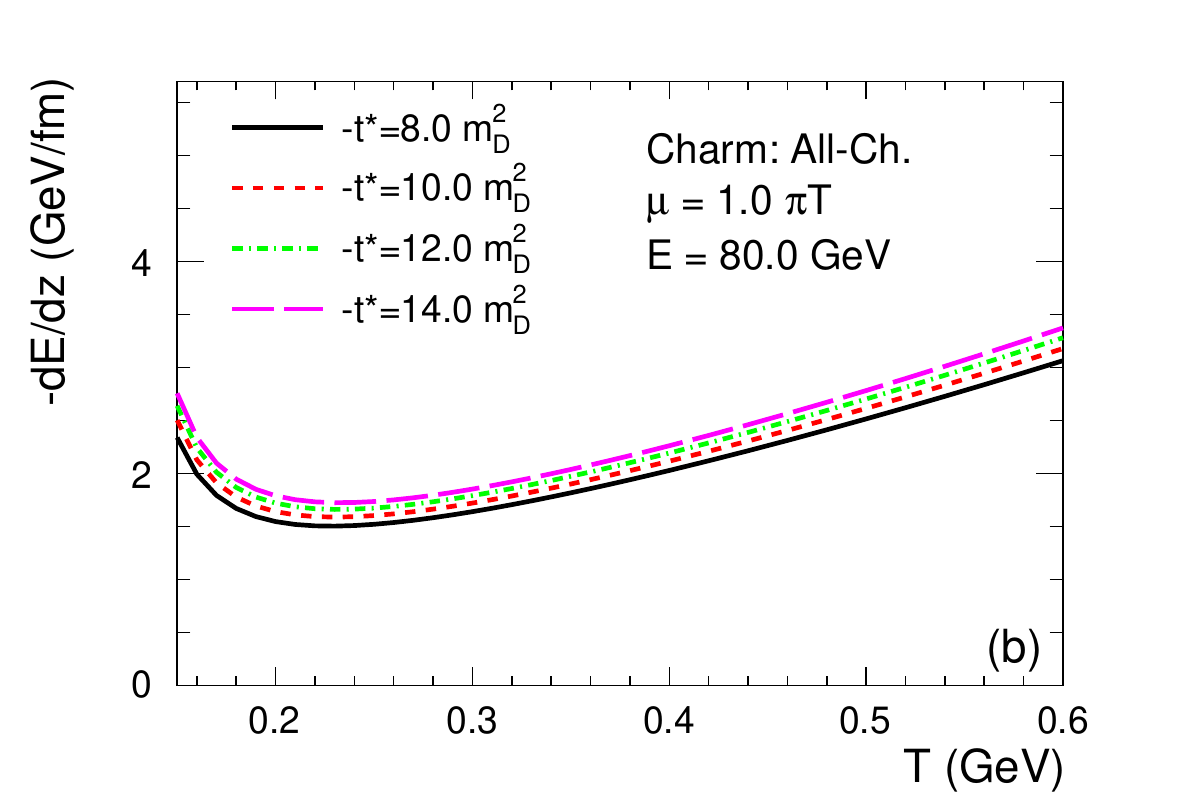}
\includegraphics[width=.47\textwidth]{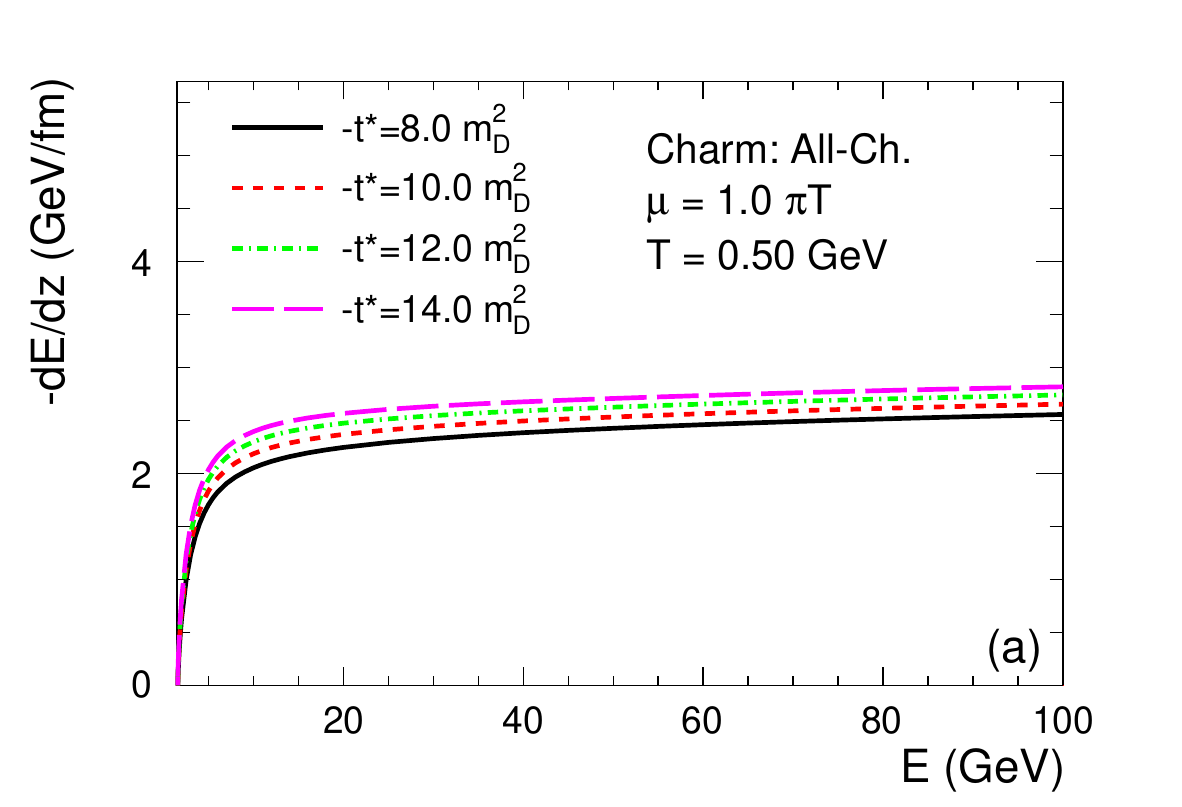}
\caption{(Color online) Comparison of the total energy loss $dE/dz$ of a charm quark
as a function of (a) temperature and (b) its energy,
displaying separately the results based on various testing parameters:
$-t^{\ast}/m_{D}^{2}=8,10,12,14$ and $\mu=\pi T$.}
\label{fig:Charm_dEdz_AllChan_variousCutOff}
\end{center}
\end{figure*}
In Fig.~\ref{fig:Charm_dEdz_AllChan_variousCutOff}, the total energy loss of charm quark is calculated,
with $-t^{\ast}/m_{D}^{2}=8,10,12,14$ and $\mu=\pi T$,
at fixed energy $E=80~{\rm GeV}$ (panel-a) and fixed temperature $T=0.5~{\rm GeV}$ (panel-b).
The maximum deviation among them is $\sim 8\%$ ($\sim 12\%$) at $T=0.16~{\rm GeV}$ ($E=2~{\rm GeV}$),
and then decreases up to $\sim4.5\%$ ($\sim6\%$) at $T=0.6~{\rm GeV}$ ($E=80~{\rm GeV}$).
Therefore, very similar to the situation of momentum diffusion coefficients~\cite{POWLANGEPJC11,Li_2021},
the energy loss also behave with a mild sensitivity to the intermediate cutoff scale $t^{\ast}$,
supporting the validity of the soft-hard approach when the coupling is not terribly small (Eq.~\ref{eq:CutOff_IO}).
Same conclusions can be drawn for bottom quarks.
Note that, with large momentum exchange, the $t^{\ast}$ dependence of $dE/dz$ vanishes
in the high-energy approximation. See Eq.~\ref{eq:dEdx_Soft_Hard_HEA} for details.


\subsection{The energy loss in the high-energy approximation}
\begin{figure*}[!htbp]
\begin{center}
\setlength{\abovecaptionskip}{-0.1mm}
\includegraphics[width=.47\textwidth]{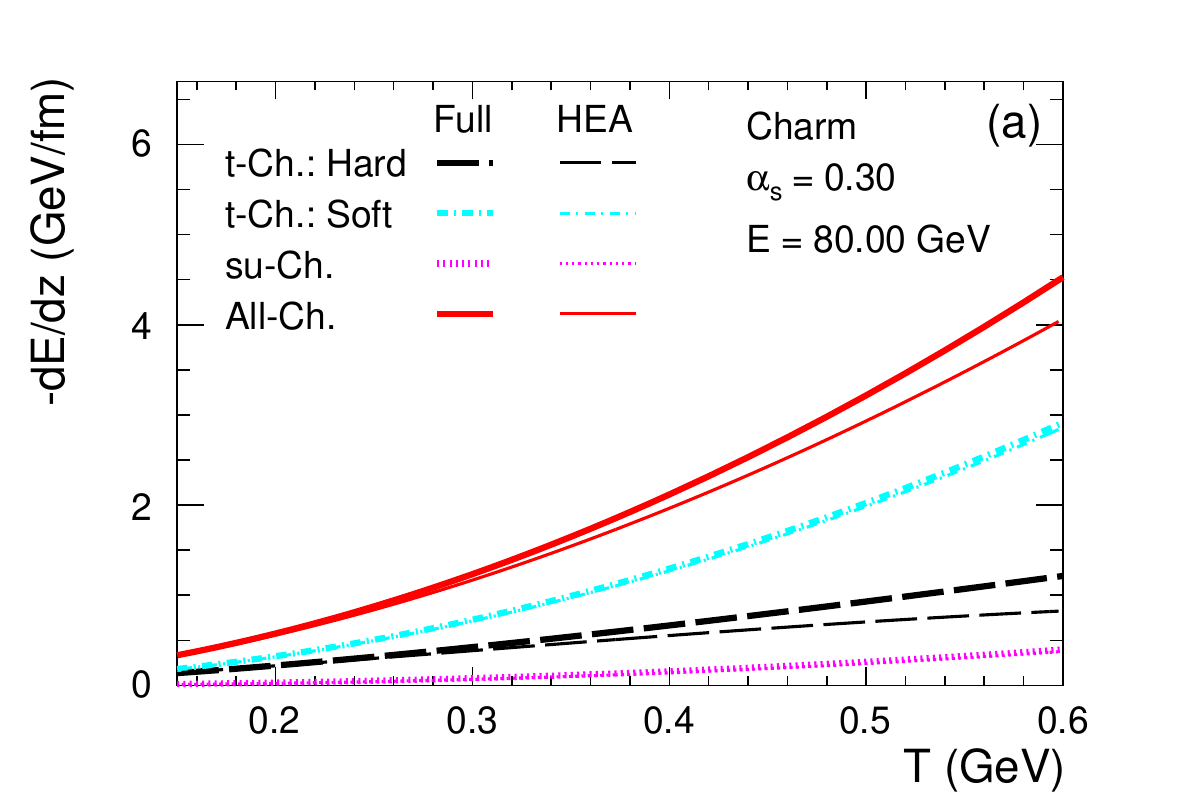}
\includegraphics[width=.47\textwidth]{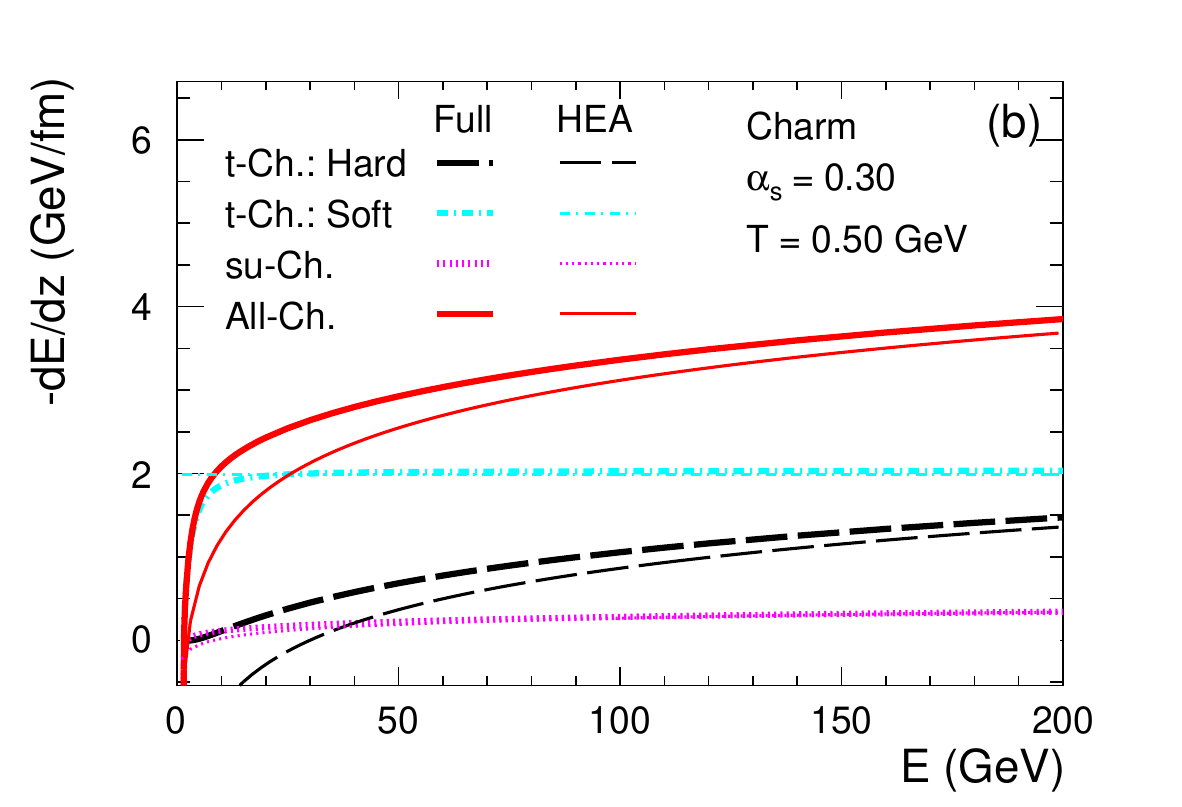}
\caption{(Color online) Comparison of charm quark energy loss based on
the full calculations (``Full''; thick curves; Eqs.~\ref{eq:dEdz_Soft}, \ref{eq:dEdz_Hard} and \ref{eq:dEdz_su})
and the high-energy approximation (``HEA'';
thin curves; Eqs.~\ref{eq:dEdx_Soft_HEA}-\ref{eq:dEdx_Soft_Hard_HEA}) at fix coupling  $\alpha_{s}=0.3$
as a function of (a) temperature and (b) energy.
The relevant results are shown as thick and thin curves, respectively, in each panel.
Various contributions to the energy loss are displayed separately as curves with different styles.
See the legend and text for details.}
\label{fig:Charm_dEdz_FullHEA}
\end{center}
\end{figure*}

In Fig.~\ref{fig:Charm_dEdz_FullHEA}(a), charm quark $dE/dz$ evaluated at fixed coupling $\alpha_{s}=0.3$ and fixed energy $E=80~{\rm GeV}$,
are presented as a function of temperature.
Same as Fig.~\ref{fig:Charm_dEdz_Full}(a),
the contributions from various channels are shown separately as curves with different styles.
We examine these calculations by systematically comparing
the results based on the full calculations (``Full''; thick curves; Eqs.~\ref{eq:dEdz_Soft}, \ref{eq:dEdz_Hard} and \ref{eq:dEdz_su})
and the high-energy approximation (``HEA''; thin curves; Eqs.~\ref{eq:dEdx_Soft_HEA}-\ref{eq:dEdx_Soft_Hard_HEA}).
It is found that:
(1) the difference between them is quite small ($\sim 2\%$)
for the contributions from the soft interaction (dot-dashed cyan curves)
in the entire region of temperature.
This is because the given energy $E=80~{\rm GeV}$,
satisfying the requirement in the ``HEA'', $E\gg m^{2}/T$ (Eq.~\ref{eq:App_sVar_HEA}),
and thus ``Full~$\approx$~HEA'' is valid in this region.
We have checked the results at low energy $E=4~{\rm GeV}$,
and sizeable difference ($\sim 30\%$) is found;
(2) for the hard interactions in $t$-channel (long dashed black curves),
a negligible difference ($\sim 5\%$ at maximum) found at low temperature $T\lesssim0.25~{\rm GeV}$,
while a visible discrepancy observed at larger temperature ($\sim 50\%$ at $T=0.6~{\rm GeV}$).
It is contributed by the scatterings off both quarks and gluons,
while the latter is dominated, in particular at high temperature,
since an additional assumption, i.e. very hard momentum exchange (Eq.~\ref{eq:App_stuRange_IO}; or backward scattering),
is adopted to obtain the corresponding results in the ``HEA'' (see Eqs.~\ref{eq:App_InteOvert_Qg_t_tmp} and \ref{eq:App_InteOvert_Qg_t});
(3) a monotonously rising temperature dependence is found
for the soft contributions obtained with a fixed coupling constant,
see Fig.~\ref{fig:Charm_dEdz_FullHEA}(a),
while a non-monotonic behavior is observed for the running coupling with temperature,
see Fig.~\ref{fig:Charm_dEdz_Full}(a);
Similar case for $s$- and $u$-channels, even though the related results are limited.

Figure~\ref{fig:Charm_dEdz_FullHEA}(b) shows the results at fixed temperature $T=0.5~{\rm GeV}$.
It is clearly observed that:
(1) the asymptotic behavior is presented towards high energy,
while a considerable variation is found at low and moderate energy for each channel;
(2) as discussed in Eq.~\ref{eq:dEdx_Soft_HEA}, the energy-dependency vanishes for the soft contribution in the ``HEA'',
which is well described by the ``Full'' calculation in asymptotically high energy region.
Therefore, the energy loss based on the `Full'' approach allows to
quantify the relevant analytical result based on the ``HEA'', in particular at high energy region.
Meanwhile, the ``Full'' calculation opens the room to study the energy loss of heavy quark
at low and moderate energy $E\lesssim50~{\rm GeV}$,
where the heavy-flavor probes are measured comprehensively at RHIC and LHC energies.

\subsection{Comparison with other models}
Here, we show the typical predictions for the collisional energy loss of heavy quarks from the other models,
and then make a comparison among them.

\begin{figure*}[!htbp]
\begin{center}
\setlength{\abovecaptionskip}{-0.1mm}
\includegraphics[width=.47\textwidth]{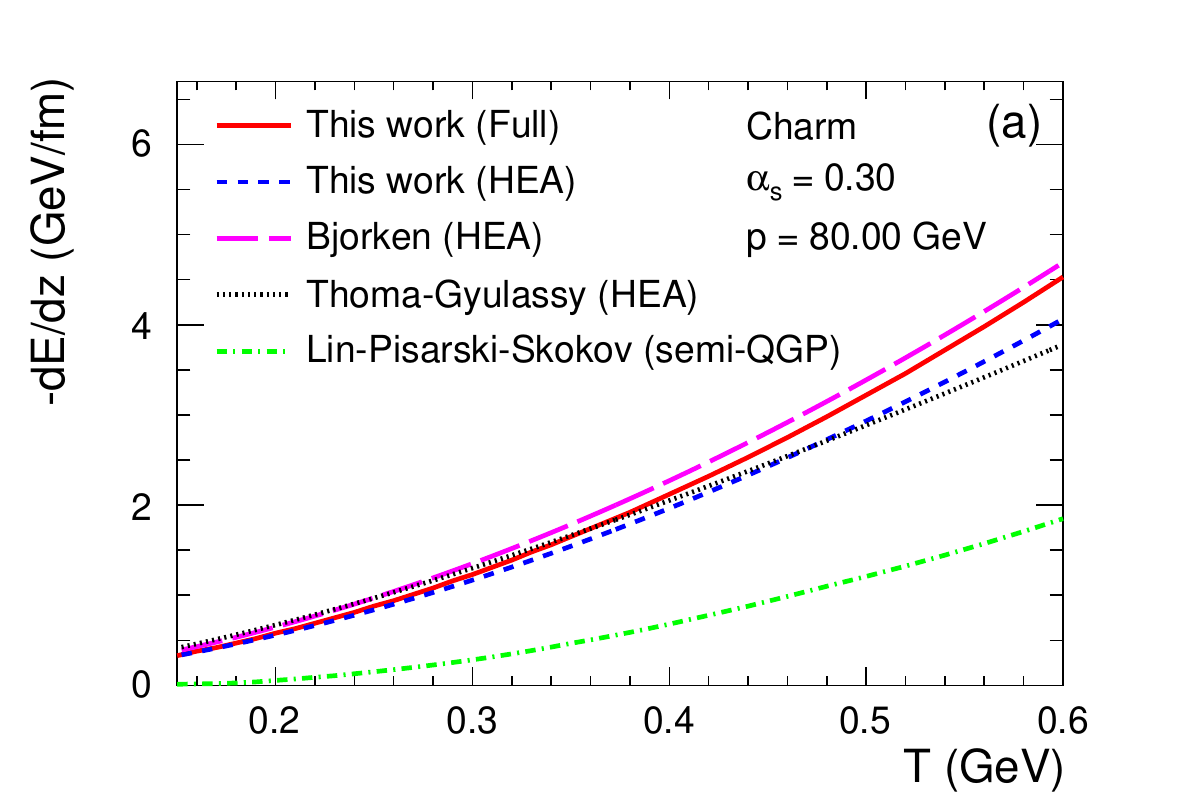}
\includegraphics[width=.47\textwidth]{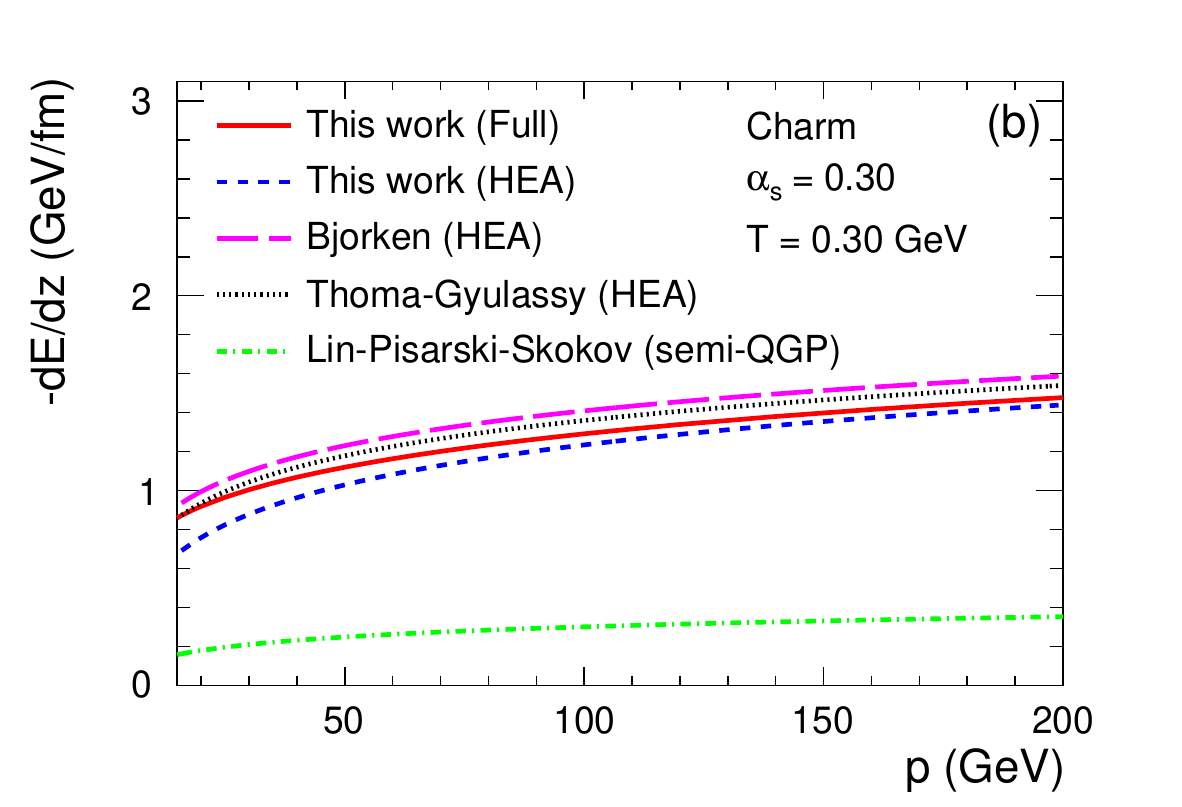}
\caption{(Color online) Left (a): the total energy loss $dE/dz$ of a charm quark as a function of temperature at fixed coupling
$\alpha_{s}=0.3$ and fixed momentum $p=80~{\rm GeV}$.
The results based on the full calculations (solid red curve; Eqs.~\ref{eq:dEdz_Soft}, \ref{eq:dEdz_Hard} and \ref{eq:dEdz_su})
and the high-energy approximation (dashed blue curve;
Eq.~\ref{eq:dEdx_Soft_Hard_HEA}) are compared to the other calculations
by Bjorken~\cite{EnergyLossMovivateJDB82} (long dashed pink curve), Thoma-Gyulassy~\cite{THOMA1991491} (dotted black curve) and
Lin-Pisarski-Skokov~\cite{LIN2014236} (dot-dashed green curve).
Right (b): same as panel-a but for $dE/dz$ as a function of heavy quark momentum at fixed temperature $T=0.3~{\rm GeV}$.}
\label{fig:Charm_dEdz_CmpModels}
\end{center}
\end{figure*}

\begin{figure*}[]
\begin{center}
\setlength{\abovecaptionskip}{-0.1mm}
\includegraphics[width=.47\textwidth]{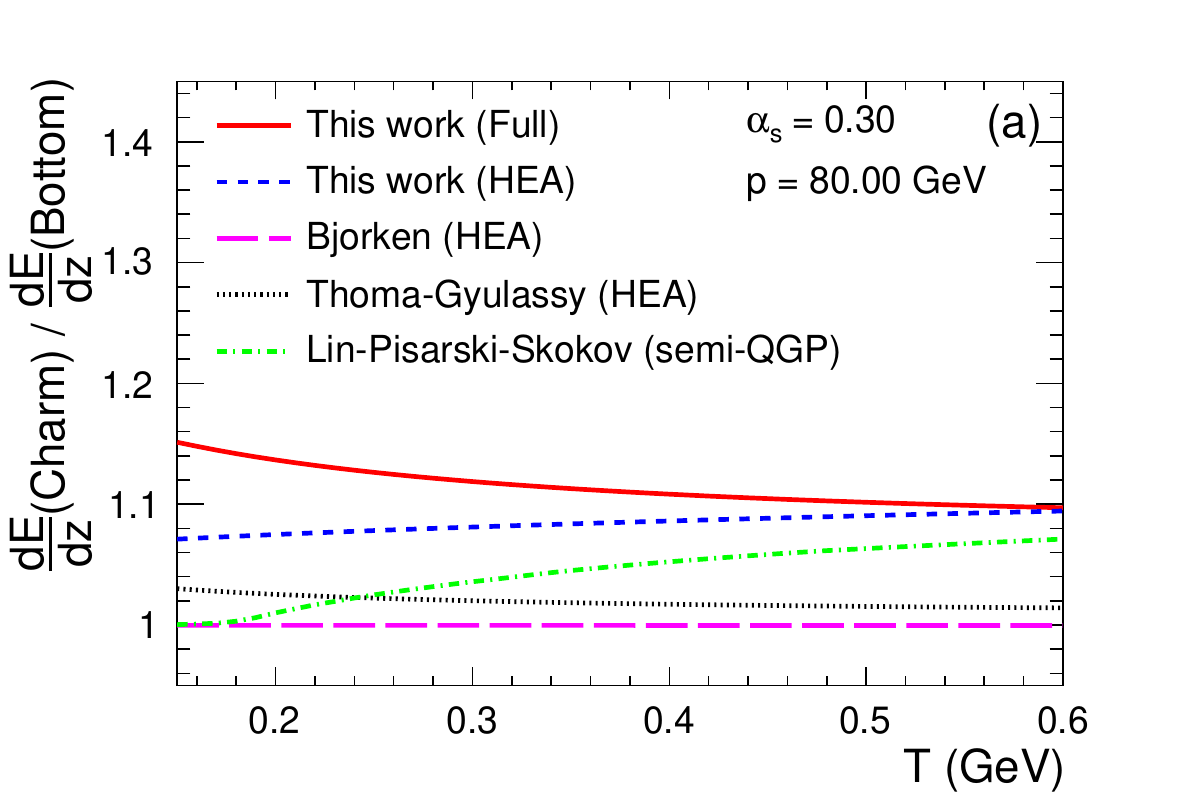}
\includegraphics[width=.47\textwidth]{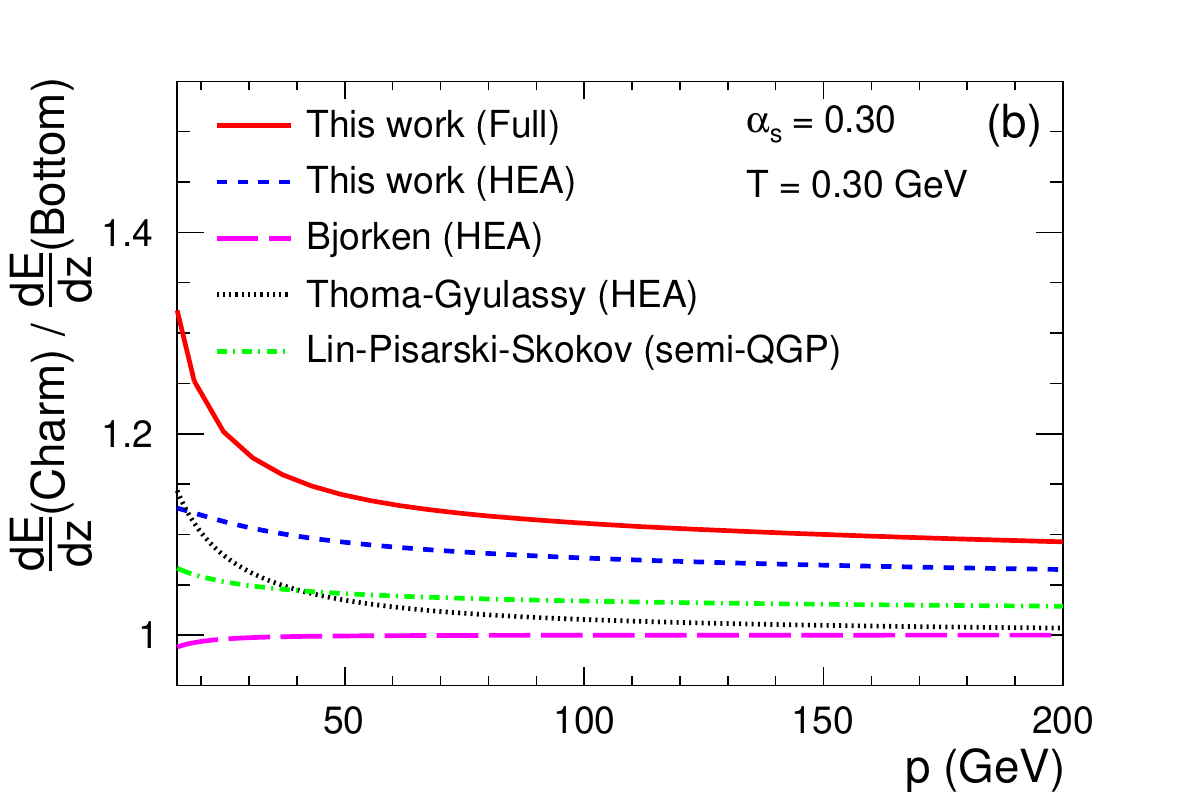}
\caption{Comparison of $dE/dz$ for charm and bottom quarks at fixed coupling $\alpha_{s}=0.3$.}
\label{fig:CharmBottomRatio_dEdz_CmpModels}
\end{center}
\end{figure*}

\textbf{\textit{Bjorken}}: In Ref.~\cite{EnergyLossMovivateJDB82}, he considered the propagation
of a massless parton with high energy through an ideal QCD medium with temperature $T$.
The resulting energy per unit length was formulated,
which can be adapted to the case of heavy quarks,
\begin{equation}
	\begin{aligned}\label{eq:Bjorken}
		-\frac{dE}{dz}=\frac{4}{3}\pi \alpha_{s}^{2}T^{2}\biggr(\frac{N_{c}}{3}+\frac{N_{f}}{6}\biggr)ln\frac{q_{max}^{2}}{q_{min}^{2}}.
	\end{aligned}
\end{equation}
To regulate the infrared and ultraviolet divergences in $t$-channels,
the invanrant four-momentum transfer $q$
is restricted within the range $-q^{2}_{min}<-q^{2}<-q^{2}_{max}$,
where, $-q_{max}^{2}=4TE_{1}$ and $-q_{min}^{2}=m_{D}^{2}$~\cite{PhysRevD.44.R2625}.
We can see that its path-dependency vanishes (see Fig.~\ref{fig:CharmBottomRatio_dEdz_CmpModels});
$dE/dz$ depends logarithmically on the quark energy;
the temperature-dependency behaves $dE/dz\propto T^{2}ln(T+const.)$ at fixed coupling.

\textbf{\textit{Thoma-Gyulassy}}: In Ref.~\cite{THOMA1991491}, Thoma and Gyulassy updated the Bjorken approach
by including a more careful treatment of the infrared divergences, yielding
\begin{equation}
	\begin{aligned}\label{eq:ThomaGyulassy}
		-\frac{dE}{dz} = &\frac{4}{3}\pi\alpha_{s}^{2}T^{2}\biggr(\frac{N_{c}}{3}+\frac{N_{f}}{6}\biggr) \\
		&\times ln\frac{4T|\vec{p}_{1}|}{(E_{1}-|\vec{p}_{1}|+4T)m_{D}^{2}},
	\end{aligned}
\end{equation}
where, $E_{1}$ ($\vec{p}_{1}$) denotes the heavy quark energy (momentum).

\textbf{\textit{Lin-Pisarski-Skokov}}: In Ref.~\cite{LIN2014236},
the authors formulated the collisional energy loss for heavy quarks
based on the semi quark–gluon plasma approach,
\begin{equation}
	\begin{aligned}\label{eq:LinShu}
		-\frac{dE}{dz} =& \pi\alpha_{s}^{2}T^{2} \biggr\{ S^{qk}\frac{N_{f}(N_{c}^{2}-1)}{12N_{c}} ln\bigg(\frac{E_{1}T}{m_{D}^{2}}\bigg) \\
		&+ S^{gl} \biggr[ \frac{N_{c}^{2}-1}{6}ln\bigg(\frac{E_{1}T}{m_{D}^{2}}\bigg) + \frac{C_{f}^{2}}{6}ln\bigg(\frac{E_{1}T}{m_{1}^{2}}\bigg) \biggr] \biggr\},
	\end{aligned}
\end{equation}
in which $C_{f}\equiv(N_{c}^{2}-1)/(2N_{c})$ indicatess the Casimir factor for the fundamental representation.
$S^{qk}$ and $S^{gl}$ indicate the suppression factor for quark and gluon, respectively, as shown in Fig.~4 of Ref.~\cite{LIN2014236}.
It is argued that the heavy quark scattering off quarks (gluons) is
suppressed by one (two) power of the Polyakov loop,
resulting in a stronger suppression of the corresponding energy loss~\cite{LIN2014236}.

Figure~\ref{fig:Charm_dEdz_CmpModels} presents the temperature (panel-a) and energy dependence (panel-b)
of $dE/dz$ as obtained, with a fixed coupling constant $\alpha_{s}=0.3$ and charm quark mass $m_{c}=1.5~{\rm GeV}$,
from the Bjorken approach~\cite{EnergyLossMovivateJDB82} (long dashed pink curves),
Thoma-Gyulassy~\cite{THOMA1991491} (dotted black curves) and Lin-Pisarski-Skokov~\cite{LIN2014236} (dot-dashed green curves),
as well as the results from the two approaches in this work,
i.e. the full calculation (solid red curves; Eqs.~\ref{eq:dEdz_Soft}, \ref{eq:dEdz_Hard} and \ref{eq:dEdz_su})
and the high-energy approximation (dashed blue curves; Eq.~\ref{eq:dEdx_Soft_Hard_HEA}).
We can see that:
(1) a common behavior, i.e. a monotonously rising temperature (momentum) dependence,
is observed for all the models,
together with the larger (smaller) gradient in the high temperature (momentum) region;
(2) concerning the temperature-dependency, except for the Lin-Pisarski-Skokov approach,
the remaining results are closer at low temperature,
while they are compatible at high temperature;
similarly, for the momentum-dependency,
they are apparently closer to each other at large momentum,
where the high-energy approximation is more reasonable;
(3) the results based on the Lin-Pisarski-Skokov approach are, as expected,
smaller than that based on the ``HEA'' approach.

To investigate the mass effect on the collisional energy loss,
we also calculate similar results for bottom quarks.
The ratio (=``Charm/Bottom'') between them is shown in Fig.~\ref{fig:CharmBottomRatio_dEdz_CmpModels}.
It is clearly shown that a charm quark ($m_{c}=1.5~{\rm GeV}$)
loses more its energy when comparing with a bottom quark ($m_{b}=4.75~{\rm GeV}$)
with the same momentum or medium temperature.
This can be explained by the fact that, to build a sizeable energy loss,
the massive heavy quark needs frequent interactions with large coupling.
Therefore, compared with charm, the bottom quark has difficulty losing its energy,
resulting in smaller $dE/dz$.
The difference between charm and bottom tends to decrease toward high momentum,
where the mass effect is expected small.
Similar behavior was found in our previous work~\cite{CTGUHybrid5}.

\subsection{The energy loss calculated with the Einstein's relation}
As discussed in Sect.~\ref{subsec:eloss_HTLpQCD} and \ref{subsec:eloss_ER},
the heavy quark energy loss, $-dE/dz=\eta_{D}|\vec{p}^{\;}|$, can be evaluated in two different scenarios:
(I) $\eta_{D}$ calculated via the scattering rate (Eq.~\ref{eq:ELoss_Def}),
\begin{equation}
	\begin{aligned}\label{eq:EtaD_ViaGamma}
		\eta_{D} = \frac{1}{|\vec{p}_{1}|} \int d^{3}\vec{q} \; \frac{d\Gamma}{d^{3}\vec{q}} \; \frac{\omega}{v_{1}};
	\end{aligned}
\end{equation}
(II) $\eta_{D}$ calculated by
taking the momentum diffusion coefficients as fundamental $\eta_{D}(\kappa_{T},\kappa_{L})$,
i.e. via the Einstein's relationship (Eqs.~\ref{eq:ELoss_EtaD_pQCDHTL}-\ref{eq:KappaL}).
In the following, these two scenarios are denoted as ``Full'' and ``Full-ER'',
indicating the energy loss without and with imposing the Einstein's relationship, respectively.

\begin{figure*}[!htbp]
\begin{center}
\setlength{\abovecaptionskip}{-0.1mm}
\includegraphics[width=.47\textwidth]{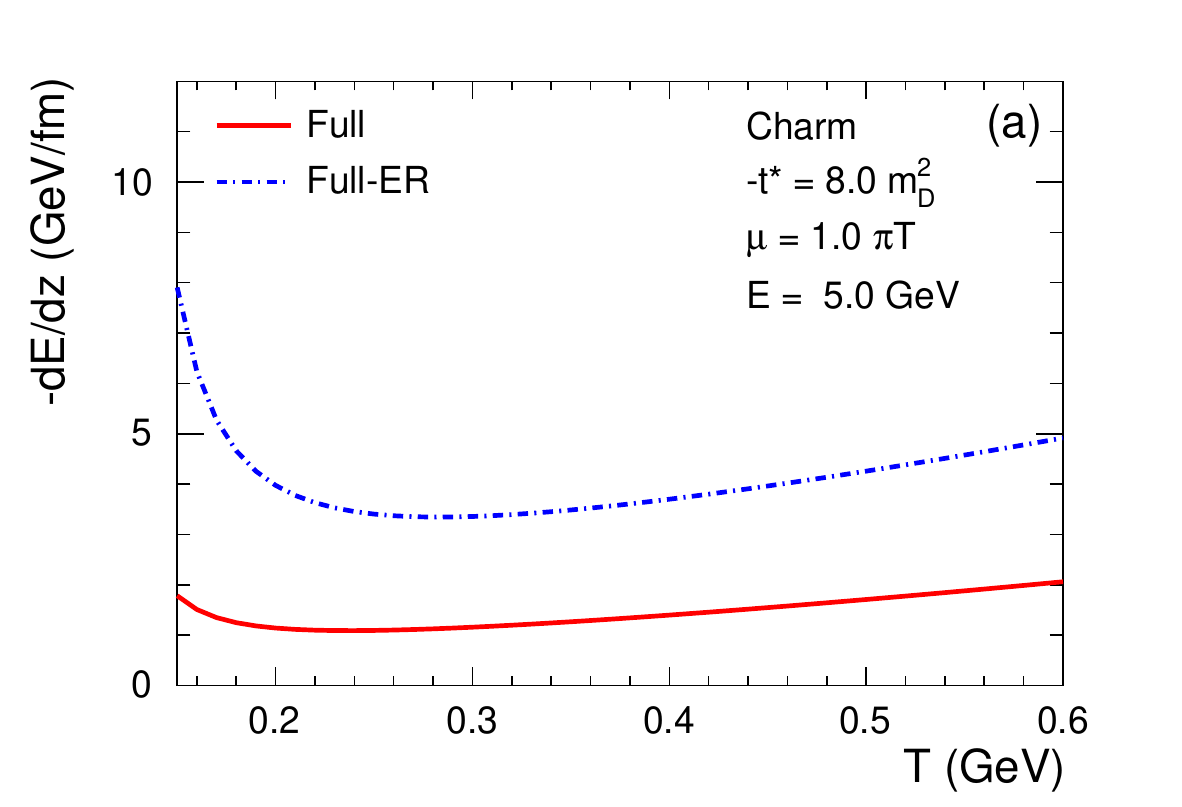}
\includegraphics[width=.47\textwidth]{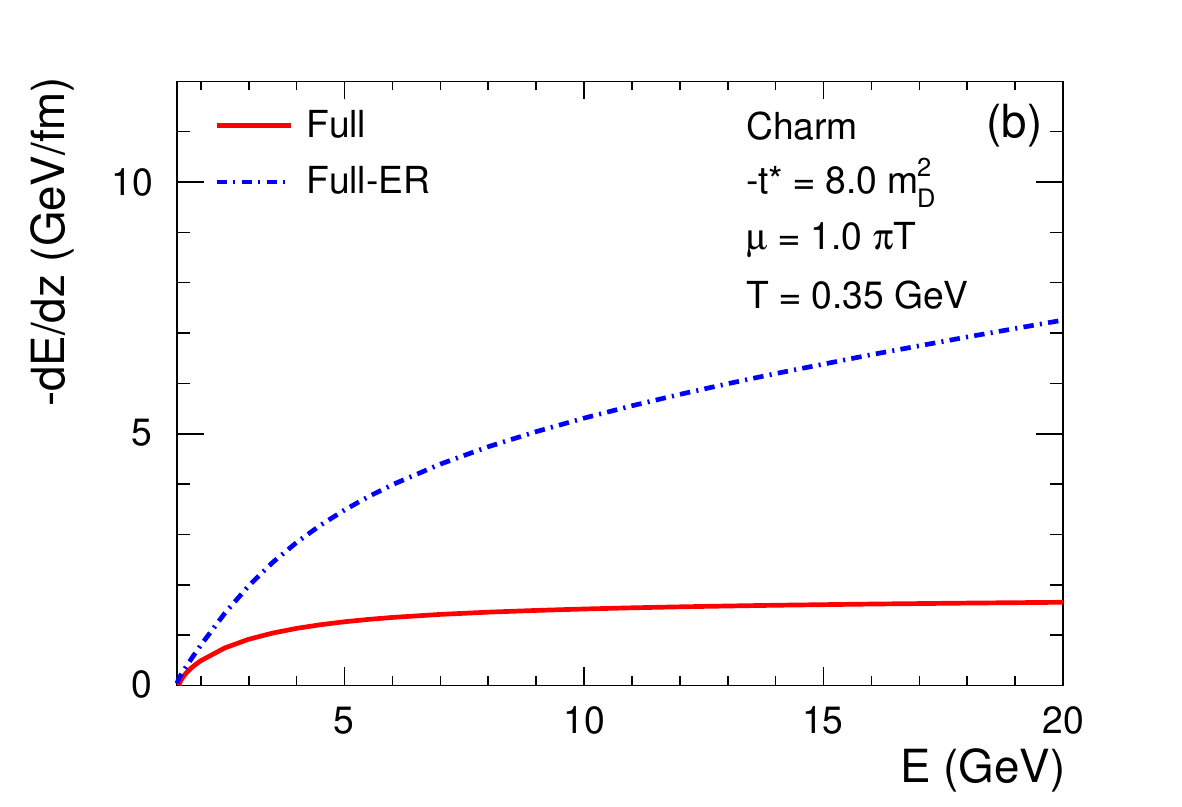}
\caption{(Color online) Energy loss of charm quarks as a function of (a) temperature and (b) energy:
results obtained with $-t^{\ast}=8m_{D}^{2}$ and $\mu=\pi T$
via the full calculations (``Full'', Eqs.~\ref{eq:dEdz_Soft}, \ref{eq:dEdz_Hard} and \ref{eq:dEdz_su})
and the Einstein's relationship (``ER'', Eqs.~\ref{eq:ELoss_EtaD_pQCDHTL}-\ref{eq:KappaL})
are shown separately as solid read and dot-dashed blue curves, respectively, in each panel.
See the text for details.}
\label{fig:Charm_dEdz_FullEinstein}
\end{center}
\end{figure*}

The energy loss of charm quark with Einstein's relationship imposed
are presented as a function of temperature (panel-a) and energy (panel-b)
in Fig.~\ref{fig:Charm_dEdz_FullEinstein} (dot-dashed blue curves).
The results without imposing Einstein's relationship
(solid red curves) are shown as well for comparison.
It is found that: (1) they show a qualitatively similar
trend but with stronger temperature and energy dependence for the former one;
(2) the former one is systematically larger than the latter one,
indicating a stronger interaction strength between the heavy quarks and the thermal medium constituents.

We can therefore expect that, when imposing the Einstein's relationship,
the energic heavy quarks will lose more their initial energy during scatterings and
approach the thermal equilibrium faster, resulting in
(1) a steeper momentum spectra, as observed in Figs.~5 and 6 of Ref.~\cite{PhysRevC.90.044901},
since stronger interactions are more powerful to pull the heavy quarks from high momentum and low momentum;
(2) a stronger suppression in the nuclear modification factor,
which is sensitive to the effects such as in-medium energy loss at moderate and high transverse momentum ($\pt\gtrsim m_{\rm Q}$);
(3) a larger elliptic flow coefficient,
which is sensitive to the path-length dependence of the in-medium
energy loss at high transverse momentum ($\pt\gg m_{\rm Q}$).
Similar behaviors were found and shown in Figs.~13 and 14 of Ref.~\cite{XuCoefficient18}.

As pointed in Ref.~\cite{XuCoefficient18},
there is an ambiguity when imposing the Einstein's relationship (Eq.~\ref{eq:EtaD_DissFluc})
since only two variables among $\eta_{D}$, $\kappa_{T}$, $\kappa_{L}$ are required in the implementation.
The resulting heavy quark energy loss, as shown in Fig.~\ref{fig:Charm_dEdz_FullEinstein},
and the relevant final observables, such as the nuclear modification factor and the elliptic flow coefficient~\cite{XuCoefficient18},
are strongly influenced by this choice.
Thus, as discussed in Ref.~\cite{XuCoefficient18,GUBSER2008175},
the Fokker-Planck and Langevin approach may not capture the whole evolution properties of heavy quarks.



\section{Summary}\label{sec:summary}
In summary we have reconsidered the heavy quark energy loss $dE/dz$ in a thermalized QCD medium.
The contributions from the binary interactions with light quarks and gluons
are formulated at leading order in QCD coupling constant,
by utilizing a recently developed soft-hard factorization approach.
It is found that the full $dE/dz$ has a mild sensitivity to the intermediate scale $t^{\ast}$,
supporting the validity of the soft-hard model when the coupling is not terribly small.
The soft contribution ($-t<-t^{\ast}$) behaves a non-monotonic dependence on the medium temperature,
which is mainly induced by the running of the coupling with temperature.
The hard contributions ($-t>-t^{\ast}$) from the $t$-channel increase with temperature.
A similar trend is observed in $s$- and $u$-channels,
even though the related results are limited comparing with the results from $t$-channel.
Due to the relativistic effect,
the soft contribution has a stronger energy dependence at low energy $E\lesssim 3~{\rm GeV}$,
while an almost flat behavior at much higher energy $E\gtrsim 10~{\rm GeV}$.

The energy loss based on the full calculations can be
simplified to an analytical form in the limit of high quark energy, $E_{1}\gg m^{2}_{1}/T$,
where $E_{1}$ is the injected heavy quark energy and $m_{1}$ is its mass.
The logarithmic contributions are obtained for the soft ($dE/dz\propto ln(-t^{\ast}/m^{2}_{D})$)
and hard interactions ($dE/dz\propto ln[E_{1}T/(-t^{\ast})]$) in $t$-channel,
together with the overall interactions ($dE/dz\propto ln(E_{1}T/m^{2}_{1})$) in $s$- and $u$-channels.
Combining all these processes, the final result cancels the $t^{\ast}$-dependence.
The results based on this high-energy approximation are compared with other model predictions.
They are similar at fixed coupling.
The energy loss for charm quark ($m_{c}=1.5~{\rm GeV}$) is systematically larger than that for bottom quark ($m_{b}=4.75~{\rm GeV}$),
in particular at low energy.
The results based on the full calculations are consistent with the analytical result at very high energy ,
while a sizeable discrepancy is noticed at low energy.
Thus, our calculations are also important to study the energy loss effect
at low and moderate energy $E\lesssim50~{\rm GeV}$,
where the heavy-flavor probes are measured comprehensively at RHIC and LHC energies.

Finally, we calculate the energy loss from the drag coefficient, $-dE/dz = \eta_{D}|\vec{p}_{1}|$,
by considering the momentum diffusion coefficients as fundamental,
i.e. $\eta_{D}=\eta_{D}(\kappa_{T/L})$, namely Einstein's relationship.
It is realized that the corresponding results are systematically larger
than the ones without imposing the Einstein's relationship.

\begin{acknowledgements}
The authors are grateful to Prof. Jinfeng Liao for helpful discussions and communications.
We thank Prof. Shu Lin for providing the inputs as shown in Fig.~\ref{fig:Charm_dEdz_CmpModels}.
This work is supported by the National Science Foundation of China (NSFC) under Grant Nos.12375137, 12005114 and 11847014.
\end{acknowledgements}


\appendix


\section{Derivation of the collisional energy loss in the soft-hard factorized approach}\label{appendix:Cal_ELoss}
\setcounter{equation}{0}
\renewcommand\theequation{A\arabic{equation}}

The energy loss of heavy quark (HQ) in soft collisions can be obtained by inserting Eq.~\ref{eq:Gamma_Soft} into Eq.~\ref{eq:ELoss_Def}.
It gives
\begin{equation}
	\begin{aligned}\label{eq:App_dEdx_Soft}
		\biggr[-\frac{dE}{dz}\biggr]^{soft}_{(t)} = &\frac{C_{F}g^{2}}{v_{1}} \int_{q} \int d\omega  \;\omega\; \bar{n}_{B}(\omega) \delta(\omega-\vec{v}_{1}\cdot\vec{q}\;) \\
		&\biggr\{ \rho_{L}(\omega,q) + \vec{v}_{1}^{\;2} \bigr[ 1-(\hat{v}_{1}\cdot\hat{q})^{2} \bigr]\rho_{T}(\omega,q) \biggr\},
	\end{aligned}
\end{equation}
in which we use the short notation
\begin{equation}
	\begin{aligned}\label{eq:App_Int_Short}
		\int_{q}... \equiv \int\frac{d^{3}\vec{q}}{(2\pi)^{3}}...
	\end{aligned}
\end{equation}
and show the thermal distribution function of the bosons ($B$) and fermions ($F$) as
\begin{equation}
	\begin{aligned}\label{eq:App_ThermalDis}
		&n_{B/F}(E) = (e^{E/T}\mp1)^{-1} \\
		&\bar{n}_{B/F} \equiv 1 \pm n_{B/F}.
	\end{aligned}
\end{equation}
$\vec{v}_{1}=\vec{p}_{1}/E_{1}$ indicates the HQ velocity,
while $\hat{v}_{1}=\vec{v}_{1}/|\vec{v}_{1}|$ denotes the unit vector in $\vec{v}_{1}$ direction.
$\theta$ is the angle between $\vec{v}_{1}$ and $\vec{q}$, thus,
the $\delta$ function in Eq.\ref{eq:App_dEdx_Soft} can be rewritten as
\begin{equation}
        \begin{aligned}\label{eq:App_DeltaFunc}
                \delta(\omega-\vec{v}_{1}\cdot\vec{q}\;) = \frac{1}{v_{1}q}\delta(cos\theta-\frac{\omega}{v_{1}q}).
        \end{aligned}
\end{equation}
Inserting Eq.~\ref{eq:App_DeltaFunc} into Eq.\ref{eq:App_dEdx_Soft} and performing the integral over the azimuthal and polar angles of $\vec{q}$,
we have
\begin{equation}
	\begin{aligned}\label{eq:App_dEdz_Soft_vsOmegaQ}
		\biggr[-\frac{dE}{dz}\biggr]^{soft}_{(t)} =& \frac{C_{F}g^{2}}{4\pi^{2}v_{1}^{2}} \int_{0}^{q_{max}} dq\;q \int_{-v_{1}q}^{v_{1}^{}q}d\omega\;\omega \bar{n}_{B}(\omega) \\
		& \biggr[\rho_{L}(\omega,q)+v_{1}^{2}(1-\frac{\omega^{2}}{v^{2}_{1}q^{2}})\rho_{T}(\omega,q) \biggr],
	\end{aligned}
\end{equation}
where, $q_{max}$ is the maximum momentum transfer in a collision of HQ with a medium parton.
Both the transverse and longitudinal spectral functions are odd (see Eq.~\ref{eq:GammaRhoT_Soft1} and \ref{eq:GammaRhoL_Soft1}),
the integral over the energy transfer $\omega$ in Eq.~\ref{eq:App_dEdz_Soft_vsOmegaQ} can be therefore expressed as
\begin{equation}
        \begin{aligned}\label{eq:App_InteOverOmega}
                &\int_{-v_{1}q}^{v_{1}^{}q}d\omega \omega \bar{n}_{B}
                \bigr[\rho_{L}+v_{1}^{2}(1-\frac{\omega^{2}}{v^{2}_{1}q^{2}})\rho_{T} \bigr] \\
		&=\int_{0}^{v_{1}q}d\omega \omega \bigr[\bar{n}_{B}(\omega)+\bar{n}_{B}(-\omega)\bigr] \bigr[ \rho_{L} + (v_{1}^{2}-\frac{\omega^{2}}{q^{2}})\rho_{T} \bigr] \\
		&=\int_{0}^{v_{1}q}d\omega \omega \bigr[ \rho_{L}(\omega,q) + (v_{1}^{2}-\frac{\omega^{2}}{q^{2}})\rho_{T}(\omega,q) \bigr]
        \end{aligned}
\end{equation}
with the identity
\begin{equation}
	\begin{aligned}
		\bar{n}_{B}(\omega)+\bar{n}_{B}(-\omega) = 1
	\end{aligned}
\end{equation}
used in the second equality in Eq.~\ref{eq:App_InteOverOmega}.

For convenience we change the variables in multiple integrals in Eq.~\ref{eq:App_dEdz_Soft_vsOmegaQ} as
\begin{equation}
	\begin{aligned}\label{eq:App_NewVariables}
		&t=\omega^{2}-q^{2}; \qquad &x=\frac{\omega}{q},
	\end{aligned}
\end{equation}
yielding
\begin{equation}
	\begin{aligned}\label{eq:App_dtdomega2dtdx}
		dtdx=\begin{vmatrix}
			\frac{\partial t}{\partial \omega} & \frac{\partial t}{\partial q} \\
			\frac{\partial x}{\partial \omega} & \frac{\partial x}{\partial q} \\
		\end{vmatrix} dqd\omega 
	= 2(1-x^{2})dqd\omega.
	\end{aligned}
\end{equation}
Substituting Eq.~\ref{eq:App_InteOverOmega}, \ref{eq:App_NewVariables} and \ref{eq:App_dtdomega2dtdx} back into Eq.~\ref{eq:App_dEdz_Soft_vsOmegaQ},
we arrive at Eq.~\ref{eq:dEdz_Soft}
\begin{equation}
	\begin{aligned}\label{eq:App_dEdz_Soft_vsTX}
		\biggr[-\frac{dE}{dz}\biggr]^{soft}_{(t)} =& \frac{C_{F}g^{2}}{8\pi^{2}v^{2}_{1}} \int^{0}_{t^{*}} dt \; (-t) \int_{0}^{v_{1}}dx \frac{x}{(1-x^{2})^{2}} \\
		& \biggr[ \rho_{L}(t,x) + (v_{1}^{2}-x^{2})\rho_{T}(t,x) \biggr].
	\end{aligned}
\end{equation}

In hard collisions of heavy quark ($Q$) and medium partons ($i=q,g$),
the relevant energy loss from $t$-channel
can be obtained by inserting Eq.~\ref{eq:Gamma_Hard} into Eq.~\ref{eq:ELoss_Def}, yielding
\begin{equation}
        \begin{aligned}\label{eq:App_ELoss_hard1}
                \biggr[-\frac{dE}{dz}\biggr]^{hard}_{Qi(t)} =& \frac{1}{2|\vec{p}_{1}|} \int_{p_{2}} \frac{n(E_{2})}{2E_{2}} \int_{p_{3}} \frac{E_{1}-E_{3}}{2E_{3}} \int_{p_{4}} \frac{\bar{n}(E_{4})}{2E_{4}} \\
		&\overline{|\mathcal{M}^{2}|}_{Qi(t)} (2\pi)^{4} \delta^{(4)}(p_{1}+p_{2}-p_{3}-p_{4})
        \end{aligned}
\end{equation}
in which $n$ and $\bar{n}$ are the thermal distributions (see Eq.~\ref{eq:App_ThermalDis}).
With the help of $\delta$-function, we can reduce the integral in Eq.~\ref{eq:App_ELoss_hard1}
down to 3-dimension in the numerical calculations, as implemented in Ref.~\cite{PhysRevD.77.114017, Li_2021}.
It gives
\begin{equation}
        \begin{aligned}\label{eq:App_dEdz_Hard_Qit}
                \biggr[-\frac{dE}{dz}\biggr]^{hard}_{Qi(t)}
                =& \frac{1}{256\pi^{3}\vec{p}_{1}^{\;2}} \int_{|\vec{p}_{2}|_{min}}^{\infty}d|\vec{p}_{2}| ~E_{2} n(E_{2}) \\
		&\int_{-1}^{cos\psi|_{max}} d(cos\psi) \int_{t_{min}}^{t^{*}}dt
                \frac{b}{a^{3}} \; \overline{|\mathcal{M}^{2}|}_{Qi(t)},
        \end{aligned}
\end{equation}
where, $\psi$ is the polar angle of $\vec{p}_{2}$.
Adding up the contributions from the quark ($i=q$) and gluon ($i=g$) from $t$-channel,
we arrive at Eq.~\ref{eq:dEdz_Hard}
\begin{equation}
	\begin{aligned}\label{eq:App_dEdz_Hard}
		\biggr[-\frac{dE}{dz}\biggr]^{hard}_{(t)} = &\sum_{i=q,g}\biggr[-\frac{dE}{dz}\biggr]^{hard}_{Qi(t)} \\
		=&\frac{1}{256\pi^{3}\vec{p}_{1}^{\;2}} \sum_{i=q,g} \int_{|\vec{p}_{2}|_{min}}^{\infty}d|\vec{p}_{2}| ~E_{2} n(E_{2}) \\
		&\int_{-1}^{cos\psi|_{max}} d(cos\psi) \int_{t_{min}}^{t^{*}}dt
		\frac{b}{a^{3}} \; \overline{|\mathcal{M}^{2}|}_{Qi(t)}.
	\end{aligned}
\end{equation}
The relevant boundaries of the integrals together with the additional notations are summarized below~\cite{Li_2021}:
\begin{equation}
	\begin{aligned}\label{eq:App_P2Min}
		|\vec{p}_{2}|_{min}=\frac{ |t^{\ast}|+\sqrt{t^{\ast 2} + 4m_{1}^{2} |t^{\ast}|} }{4(E_{1}+|\vec{p}_{1}|)},
	\end{aligned}
\end{equation}
\begin{equation}
	\begin{aligned}\label{eq:App_PhiMax}
		cos\psi|_{max}=min \biggr\{ 1, \frac{E_{1}}{|\vec{p}_{1}|} - \frac{|t^{\ast}| + \sqrt{t^{\ast 2} +
		4m_{1}^{2} |t^{\ast}|}}{4|\vec{p}_{1}| \cdot |\vec{p}_{2}|} \biggr\},
	\end{aligned}
\end{equation}
\begin{equation}
	\begin{aligned}\label{eq:App_tMin}
		t_{min}=-\frac{(s-m_{1}^{2})^{2}}{s},
	\end{aligned}
\end{equation}
\begin{equation}
	\begin{aligned}\label{eq:App_aVal}
		a= \frac{s-m_{1}^{2}}{|\vec{p}_{1}|},
	\end{aligned}
\end{equation}
\begin{equation}
	\begin{aligned}\label{eq:App_bVal}
		b=-\frac{2t}{\vec{p}_{1}^{\;2}} \bigr[ E_{1}(s-m_{1}^{2})-E_{2}(s+m_{1}^{2}) \bigr].
	\end{aligned}
\end{equation}
\begin{equation}
        \begin{aligned}\label{eq:App_cVal}
        c=-\frac{t}{\vec{p}_{1}^{\;2}} \biggr\{ t\bigr[ (E_{1}+E_{2})^{2}-s \bigr] +
	4\vec{p}_{1}^{\;2}\vec{p}_{2}^{\;2}sin^{2}\psi \biggr\}.
        \end{aligned}
\end{equation}
\begin{equation}
        \begin{aligned}\label{eq:App_DVal}
        D \equiv b^{2}+4a^{2}c=-t \biggr[ ts + (s-m_{1}^{2})^{2} \biggr] \cdot \biggr( \frac{4E_{2}sin\psi}{|\vec{p}_{1}|} \biggr)^{2}.
        \end{aligned}
\end{equation}

The vacuum matrix elements $\overline{|\mathcal{M}^{2}|}_{Qi}$ for quark ($i=q$) and gluon ($i=g$) are expressed, respectively, as~\cite{Combridge79, Li_2021}
\begin{equation}
        \begin{aligned}\label{eq:App_MatrixHQq}
                &\overline{|\mathcal{M}^{2}|}_{Qq(t)} =
		\frac{16}{9} N_{f} N_{c} g^{4} \biggr[ \frac{2\tilde{s}^{\;2}}{t^{2}}+\frac{2(\tilde{s}+m_{1}^{2})}{t}+1 \biggr],
        \end{aligned}
\end{equation}
\begin{equation}
        \begin{aligned}\label{eq:App_MatrixHQg_t}
                \overline{|\mathcal{M}^{2}|}_{Qg(t)} =& 2(N_{c}^{2}-1) g^{4} \biggr[ -\frac{2\tilde{s}\tilde{u}}{t^{2}} +\frac{m_{1}^{2}(\tilde{s}-\tilde{u})-\tilde{s}\tilde{u}}{t\tilde{s}} \\
		&- \frac{m_{1}^{2}(\tilde{s}-\tilde{u})+\tilde{s}\tilde{u}}{t\tilde{u}} \biggr],
        \end{aligned}
\end{equation}
\begin{equation}
        \begin{aligned}\label{eq:App_MatrixHQg_su}
                \overline{|\mathcal{M}^{2}|}_{Qg(s+u)} =& \frac{8}{9} (N_{c}^{2}-1) g^{4} \biggr[ \frac{2m_{1}^{2}(\tilde{s}+2m_{1}^{2})-\tilde{s}\tilde{u}}{\tilde{s}^{\;2}} + \\
                &\frac{2m_{1}^{2}(\tilde{u}+2m_{1}^{2})-\tilde{s}\tilde{u}}{\tilde{u}^{\;2}} - \frac{m_{1}^{2}(4m_{1}^{2}-t)}{4\tilde{s}\tilde{u}} \biggr].
        \end{aligned}
\end{equation}
Here, we have introduced the abbreviation,
\begin{equation}
        \begin{aligned}
\tilde{s}\equiv s-m_{1}^{2}, \qquad \tilde{u}\equiv u-m_{1}^{2},
        \end{aligned}
\end{equation}
where, $m_{1}$ denotes the mass of the injected heavy quark.
The Mandelstam relation can be rewritten as $\tilde{s}+\tilde{u}+t=0$.
For $Qg$ scattering, the contributions from the $t$-channel, $\overline{|\mathcal{M}^{2}|}_{Qg(t)}$,
and $s$- and $u$-channels, $\overline{|\mathcal{M}^{2}|}_{Qg(s+u)}$,
are shown in Eq.~\ref{eq:App_MatrixHQg_t} and \ref{eq:App_MatrixHQg_su}, respectively.

The scattering of $Qg$ in the $s$- and $u$-channels
does not give rise to an infrared divergence from the small momentum transfer $t\rightarrow 0$.
The intermediate cutoff can be therefore set to zero, $t^{\ast}=0$,
leading to $|\vec{p}_{2}|_{min}=0$ (Eq.~\ref{eq:App_P2Min}) and $cos\psi|_{max}=1$ (Eq.~\ref{eq:App_PhiMax}).
Accordingly, the energy loss from the $s$- and $u$-channels
can be obtained by modifying the relevant result from the $t$-channel (Eq.~\ref{eq:App_dEdz_Hard}),
\begin{equation}
\begin{aligned}\label{eq:App_dEdz_su}
\biggr[-\frac{dE}{dz}\biggr]_{(s+u)} =& \frac{1}{256\pi^{3}\vec{p}_{1}^{\;2}} \int_{0}^{\infty}d|\vec{p}_{2}| ~E_{2} n(E_{2}) \\
&\int_{-1}^{1} d(cos\psi) \int_{t_{min}}^{0}dt \frac{b}{a^{3}} \; \overline{|\mathcal{M}^{2}|}_{Qg(s+u)},
\end{aligned}
\end{equation}
as quoted in Eq.~\ref{eq:dEdz_su}.

\section{Derivation of the analytical results in the high-energy approximation}\label{appendix:Cal_HighEnergyLimit}
\setcounter{equation}{0}
\renewcommand\theequation{B\arabic{equation}}

In this appendix, we derive the analytical results for
the heavy quark (HQ) energy loss in the high-energy limit $E_{1}\rightarrow \infty$.
For both soft and hard contributions,
we will first investigate the kinematic constraints in this limit,
and then perform the relevant calculations for the energy loss.

The injected HQ moving with the velocity $v_{1}=|\vec{p}_{1}|/E_{1} \rightarrow 1$
in the high-energy approximation $E_{1}\rightarrow \infty$.
Thus, for soft collisions of HQ and medium partons,
where the momentum transfer is small,
the energy loss of HQ per traveling length (Eq.~\ref{eq:App_dEdz_Soft_vsTX}) can be modified as
\begin{widetext}
\begin{equation}
        \begin{aligned}\label{eq:App_dEdz_Tmp1}
	 &\biggr[-\frac{dE}{dz}\biggr]^{soft}_{(t)} \approx \frac{C_{F}g^{2}}{8\pi^{2}} \int_{0}^{1}dx \int^{0}_{t^{*}} dt \; (-t)
	  \biggr\{ \frac{x}{(1-x^{2})^{2}} \bigr[ \rho_{L}(t,x) + (1-x^{2})\rho_{T}(t,x) \bigr] \biggr\}.
        \end{aligned}
\end{equation}
With Eq.~\ref{eq:GammaRhoT_Soft2} and \ref{eq:GammaRhoL_Soft2}, it is found that,
\begin{equation}
        \begin{aligned}
                &\biggr\{ \frac{x}{(1-x^{2})^{2}} \bigr[ \rho_{L} + (v_{1}^{2}-x^{2})\rho_{T} \bigr] \biggr\}
		 = \pi m_{D}^{2} x^{2} \biggr[ (-t+\mathcal{A})^{2} + 4\mathcal{C}^{2} \biggr]^{-1}
		 + \frac{\pi m_{D}^{2}x^{2}}{2} \biggr[ (-t+\mathcal{B})^{2} + \mathcal{C}^{2} \biggr]^{-1}
        \end{aligned}
\end{equation}
in which
\begin{equation}
        \begin{aligned}\label{eq:App_Defined_1}
	\mathcal{A}(x) &\equiv m_{D}^{2} \cdot \mathcal{F}_{A}(x) = m_{D}^{2} \cdot \biggr[ (1-x^{2}) \bigr(1-\frac{x}{2}ln\frac{1+x}{1-x}\bigr) \biggr] \\
	\mathcal{B}(x) &\equiv m_{D}^{2} \cdot \mathcal{F}_{B}(x) = m_{D}^{2} \cdot \biggr[ x^{2} \bigr(1+\frac{1-x^{2}}{2x}ln\frac{1+x}{1-x}\bigr) \biggr] \\
	\mathcal{C}(x) &\equiv m_{D}^{2} \cdot \mathcal{F}_{C}(x) = m_{D}^{2} \cdot \biggr[ \frac{\pi}{4} x(1-x^{2}) \biggr].
        \end{aligned}
\end{equation}
Concerning the condition $m_{D}^{2}\ll -t^{\ast}\ll T^{2}$ (Eq.~\ref{eq:CutOff_IO}) we are interested in,
it is useful to define the variable,
\begin{equation}
        \begin{aligned}\label{eq:App_Defined_2}
        \mathcal{K} &\equiv \frac{-t^{\ast}}{m_{D}^{2}} \gg 1,
        \end{aligned}
\end{equation}
for further calculations.
In terms of the quantities defined in Eqs.~\ref{eq:App_Defined_1} and \ref{eq:App_Defined_2},
the integral over $t$ in Eq.~\ref{eq:App_dEdz_Tmp1} can be expressed as
\begin{equation}
        \begin{aligned}\label{eq:App_dEdz_Tmp2}
                &\int^{0}_{t^{*}} dt(-t) \biggr\{ \frac{x}{(1-x^{2})^{2}} \bigr[ \rho_{L}(t,x) + (v_{1}^{2}-x^{2})\rho_{T}(t,x) \bigr] \biggr\} \\
		&\approx \frac{3}{2}\pi m_{D}^{2} x^{2} ln(2\mathcal{K})
                 + \pi m_{D}^{2} \biggr\{ x^{2} \biggr[ \frac{1}{2}ln\frac{(\mathcal{K}+2\mathcal{F}_{A})^{2}+4\mathcal{F}_{C}^{2}}{16\mathcal{K}^{2}(\mathcal{F}_{A}^{2}+\mathcal{F}_{C}^{2})}
                 - \frac{2\mathcal{F}_{C}}{\mathcal{F}_{A}} \bigr(arctan\frac{\mathcal{K}+2\mathcal{F}_{A}}{2\mathcal{F}_{C}} - arctan\frac{\mathcal{F}_{A}}{\mathcal{F}_{C}} \bigr) \biggr] \\
                &\quad + \frac{x^{2}}{2} \biggr[ \frac{1}{2}ln\frac{(\mathcal{K}+2\mathcal{F}_{B})^2+\mathcal{F}_{C}^{2}}{4\mathcal{K}^{2}(4\mathcal{F}_{B}^{2}+\mathcal{F}_{C}^{2})}
                 - \frac{\mathcal{F}_{C}}{\mathcal{F}_{B}} \bigr(arctan\frac{\mathcal{K}+2\mathcal{F}_{B}}{\mathcal{F}_{C}} - arctan\frac{2\mathcal{F}_{B}}{\mathcal{F}_{C}} \bigr) \biggr] \biggr\} \\
                 &\equiv \frac{3}{2}\pi m_{D}^{2} x^{2} ln(2\mathcal{K}) + \pi m_{D}^{2}\cdot\mathcal{G}(x)
        \end{aligned}
\end{equation}
\end{widetext}
By substituting Eq.~\ref{eq:App_dEdz_Tmp2} back into Eq.~\ref{eq:App_dEdz_Tmp1}
and performing the remaining integral over $x$, we arrive at
\begin{equation}
        \begin{aligned}\label{eq:App_dEdz_Tmp3}
                \biggr[-\frac{dE}{dz}\biggr]^{soft-HEA}_{(t)} &\approx \frac{C_{F}g^{2}m_{D}^{2}}{16\pi} \biggr[ ln\bigr(\frac{-2t^{\ast}}{m_{D}^{2}}\bigr)
		+ 2\int_{0}^{1}dx~\mathcal{G}(x) \biggr].
        \end{aligned}
\end{equation}
It is difficult to evaluate analytically the integral in Eq.~\ref{eq:App_dEdz_Tmp3},
which is vanished by checking numerically.
Therefore, Eq.~\ref{eq:App_dEdz_Tmp3} can be further reduced to
\begin{equation}
        \begin{aligned}\label{eq:App_dEdx_Soft_HEA}
                \biggr[-\frac{dE}{dz}\biggr]^{soft-HEA}_{(t)} &\approx \frac{C_{F}}{16\pi} \biggr( \frac{N_{c}}{3}+\frac{N_{f}}{6} \biggr) g^{4} T^{2} ln\frac{-2t^{\ast}}{m_{D}^{2}}
        \end{aligned}
\end{equation}

For hard collisions, since the incident medium partons
are in thermal equilibrium at a temperature $T$,
and their energy read $E_{2}\sim \mathcal{O}(T)$.
In the limit of high-energy $E_{1}\rightarrow \infty$,
the Mandelstam variable $s$ (Eq.~\ref{eq:MandVar}) behaves
\begin{equation}
        \begin{aligned}
                &s=m^{2}_{1}+2E_{1}E_{2}(1-cos\theta_{\vec{p}_{1}\vec{p}_{2}}) \rightarrow \infty,
        \end{aligned}
\end{equation}
for the scattering of HQ and massless partons in the laboratory frame.
It yields
\begin{equation}
        \begin{aligned}\label{eq:App_sVar_HEA}
 		&s \sim \mathcal{O}(E_{1}T) \gg m^{2}_{1} \gg -t^{\ast}, \\
		&\tilde{s}\equiv s-m^{2}_{1} \approx s.
        \end{aligned}
\end{equation}
Considering the results for the variables $t$ and $u$,
\begin{equation}
        \begin{aligned}
                &t_{min}=-\tilde{s}+\frac{m^{2}_{1}\tilde{s}}{s} \equiv -\tilde{s} + (-\tilde{u})_{min}, \\
                &-\tilde{u}<s+m^{2}_{1}=(-\tilde{u})_{max},
        \end{aligned}
\end{equation}
we have $(-\tilde{u})_{min} < -\tilde{u} < (-\tilde{u})_{max}$, more specifically
\begin{equation}
        \begin{aligned}\label{eq:App_uVar}
                &\frac{m^{2}_{1}\tilde{s}}{s} < -\tilde{u} < s+m^{2}_{1},
        \end{aligned}
\end{equation}
which can be rewritten as
\begin{equation}
        \begin{aligned}\label{eq:App_uVar_HEA}
                &m^{2}_{1} < -\tilde{u} < \tilde{s},
        \end{aligned}
\end{equation}
in the high-energy limit (Eq.~\ref{eq:App_sVar_HEA}).

Using Eq.~\ref{eq:App_sVar_HEA}, we rewrite Eq.~\ref{eq:App_P2Min}--Eq.~\ref{eq:App_bVal} as
\begin{equation}
	\begin{aligned}\label{eq:App_MidPara_HEA}
		&|\vec{p}_{2}|_{min} \approx 0, \qquad cos\psi|_{max} \approx 1, \qquad t_{min} \approx -s, \\
		&a \approx \frac{s}{E_{1}}, \qquad b \approx -\frac{2ts}{E_{1}}.
	\end{aligned}
\end{equation}

Considering the high-energy approximation (HEA, Eq.~\ref{eq:App_sVar_HEA}),
the energy loss in $t$-channel hard scattering (Eq.~\ref{eq:App_dEdz_Hard}) can be therefore simplified as
\begin{equation}
        \begin{aligned}\label{eq:App_dEdz_Hard_HEA}
                &\biggr[-\frac{dE}{dz}\biggr]^{hard-HEA}_{Qi(t)} \\
                &= \int_{p_{2}} \frac{n(E_{2})}{2E_{2}} \biggr\{ \int^{s}_{-t^{\ast}}d(-t) ~(-t) \biggr[ \frac{d\sigma}{dt} \biggr]_{Qi(t)} \biggr\},
        \end{aligned}
\end{equation}
where, the differential corss-section is explicitly given by
\begin{equation}
        \begin{aligned}\label{eq:App_DiffSigma}
		\frac{d\sigma}{dt} = \frac{\overline{|\mathcal{M}^{2}|}}{16\pi\tilde{s}^{\;2}}.
        \end{aligned}
\end{equation}
Thus, for $Qq$ scattering (Eq.~\ref{eq:App_MatrixHQq}), the integral over $-t\in(-t^{\ast},s)$,
i.e. the terms in the curly brace on the right-hand side of Eq.~\ref{eq:App_dEdz_Hard_HEA},
is expressed as
\begin{equation}
        \begin{aligned}\label{eq:App_InteOvert_Qq_t}
		&\int^{s}_{-t^{\ast}}d(-t) ~(-t) \biggr[ \frac{d\sigma}{dt} \biggr]_{Qq(t)} \\
                &\approx \frac{N_{f} N_{c}}{9\pi} g^{4} \int^{s}_{-t^{\ast}}d(-t) \biggr( \frac{2}{-t} - \frac{2}{\tilde{s}} + \frac{-t}{\tilde{s}^{\;2}} \biggr) \\
                &\approx \frac{2N_{f} N_{c}}{9\pi} g^{4} \biggr( ln\frac{s}{-t^{\ast}} - \frac{3}{4} \biggr).
        \end{aligned}
\end{equation}
Inserting Eq.~\ref{eq:App_InteOvert_Qq_t} into Eq.~\ref{eq:App_dEdz_Hard_HEA}, and using
\begin{equation}
        \begin{aligned}\label{eq:App_InteFunc_Fermi1}
        	&\int_{0}^{\infty}dE\; E n_{F}(E) = \frac{\pi^{2}T^{2}}{12}
        \end{aligned}
\end{equation}
\begin{equation}
	\begin{aligned}\label{eq:App_InteFunc_Fermi2}
			&\int_{0}^{\infty}dE \;E n_{F}(E) ln\frac{E}{a} = \frac{\pi^{2}T^{2}}{12}\biggr[ ln\frac{2T}{a}+1-\gamma_{E}+\frac{\zeta^{'}(2)}{\zeta(2)}\biggr]
	\end{aligned}
\end{equation}
for the integral over the magnitude of $\vec{p}_{2}$, we can get
\begin{equation}
        \begin{aligned}\label{eq:App_dEdx_Hard_HEA_Qq}
                &\biggr[-\frac{dE}{dz}\biggr]^{hard-HEA}_{Qq(t)} \approx \frac{N_{f}N_{c}}{216\pi} g^{4}T^{2} \biggr( ln \frac{8E_{1}T}{-t^{\ast}}-\frac{3}{4}+c \biggr),
        \end{aligned}
\end{equation}
where, the constant factor $c\equiv-\gamma_{E}+\zeta^{'}(2)/\zeta(2) \approx -1.14718$ with the Euler constant $\gamma_{E} \approx 0.57722$
and the Riemann function $\zeta$ and its derivative $\zeta^{'}$, $\zeta^{'}(2)/\zeta(2)\approx -0.56996$.
Note that the conditions $E_{1}\gg m^{2}_{1}/T$ (Eq.~\ref{eq:App_sVar_HEA})
is utilized in Eq.~\ref{eq:App_dEdx_Hard_HEA_Qq}.

For $Qg$ scattering in $t$-channel (Eq.~\ref{eq:App_MatrixHQg_t}), the integral over $-t\in(-t^{\ast},s)$, is given by
\begin{equation}
        \begin{aligned}\label{eq:App_InteOvert_Qg_t_tmp}
                &\int^{s}_{-t^{\ast}}d(-t) ~(-t) \biggr[ \frac{d\sigma}{dt} \biggr]_{Qg(t)} \\
                &= \frac{N_{c}^{2}-1}{8\pi} g^{4} \int^{s}_{-t^{\ast}}d(-t) \biggr[ \biggr( \frac{2}{-t} - \frac{2}{\tilde{s}} + \frac{-t}{\tilde{s}^{\;2}} \biggr) \\
		&\quad -\frac{m_{1}^{2}(\tilde{s}-\tilde{u})}{\tilde{s}^{\;2}} \biggr( \frac{1}{\tilde{s}} - \frac{1}{\tilde{u}} \biggr) \biggr],
        \end{aligned}
\end{equation}
in which the contribution of the first inner bracket
is evaluated as for the scattering off quark (see the first equality in Eq.~\ref{eq:App_InteOvert_Qq_t}),
while the contribution of the second vanishes for the very hard momentum exchange,
\begin{equation}
        \begin{aligned}\label{eq:App_stuRange_IO}
                &-t \approx \tilde{s} \approx s \gg m^{2}_{1} \iff -\tilde{u} \ll \tilde{s} \approx s.
        \end{aligned}
\end{equation}
Note that, with Eq.~\ref{eq:App_stuRange_IO},
the Mandelstam invariant $t$ reads
\begin{equation}
        \begin{aligned}
                &-t = \frac{\tilde{s}^{\;2}}{2s} (1-cos\theta_{\vec{p}_{1}\vec{p}_{3}}) \approx \tilde{s},
        \end{aligned}
\end{equation}
corresponding to the backward scattering $\theta_{\vec{p}_{1}\vec{p}_{3}} = \pi$
in the center-of-mass frame of HQ and medium parton.
Therefore, Eq.~\ref{eq:App_InteOvert_Qg_t_tmp} can be simplified in this region
\begin{equation}
        \begin{aligned}\label{eq:App_InteOvert_Qg_t}
                &\int^{s}_{-t^{\ast}}d(-t) ~(-t) \biggr[ \frac{d\sigma}{dt} \biggr]_{Qg(t)} 
                \approx \frac{N_{c}^{2}-1}{4\pi} g^{4} \biggr( ln\frac{s}{-t^{\ast}} - \frac{3}{4} \biggr).
        \end{aligned}
\end{equation}
The logarithm in Eq.~\ref{eq:App_InteOvert_Qg_t} arises from the range as shown in Eq.~\ref{eq:App_stuRange_IO} in the $t$-channel,
and the constant part comes from its interference terms.
Inserting Eq.~\ref{eq:App_InteOvert_Qg_t} into Eq.~\ref{eq:App_dEdz_Hard_HEA}, and using
\begin{equation}
        \begin{aligned}\label{eq:App_InteFunc_Bose1}
               &\int_{0}^{\infty}dE\; E n_{B}(E) = \frac{\pi^{2}T^{2}}{6}
        \end{aligned}
\end{equation}
\begin{equation}
	\begin{aligned}\label{eq:App_InteFunc_Bose2}
			&\int_{0}^{\infty}dE \;E n_{B}(E) ln\frac{E}{a} = \frac{\pi^{2}T^{2}}{6}\biggr[ ln\frac{T}{a}+1-\gamma_{E}+\frac{\zeta^{'}(2)}{\zeta(2)}\biggr]
	\end{aligned}
\end{equation}
for the integral over $p_{2}$, we arrive at
\begin{equation}
        \begin{aligned}\label{eq:App_dEdx_Hard_HEA_Qg_t}
                &\biggr[-\frac{dE}{dz}\biggr]^{hard-HEA}_{Qg(t)} = \frac{N_{c}^{2}-1}{96\pi} g^{4} T^{2} \biggr( ln \frac{4E_{1}T}{-t^{\ast}}-\frac{3}{4}+c \biggr).
        \end{aligned}
\end{equation}

For $Qg$ scattering in $s$- and $u$-channels (Eq.~\ref{eq:App_MatrixHQg_su}), the integral over $-t\in(0,s)$, is characterized by
\begin{equation}
        \begin{aligned}\label{eq:App_InteOvert_Qg_su_tmp}
                &\int^{s}_{0}d(-t) ~(-t) \biggr[ \frac{d\sigma}{dt} \biggr]_{Qg(s+u)} \\
		&= \frac{N_{c}^{2}-1}{18\pi} g^{4} \int^{s}_{0}d(-t) \frac{-t}{\tilde{s}^{\;2}} \biggr\{ \biggr( -\frac{\tilde{u}}{\tilde{s}} - \frac{\tilde{s}}{\tilde{u}} \biggr) \\
                &~ +\biggr[ \frac{2m_{1}^{2}(s+m_{1}^{2})}{\tilde{s}^{\;2}} + \frac{2m_{1}^{2}(m_{1}^{2}+u)}{\tilde{u}^{\;2}} - \frac{m_{1}^{2}(4m_{1}^{2}-t)}{4\tilde{s}\tilde{u}} \biggr] \biggr\}, \\
        \end{aligned}
\end{equation}
in which the contribution of the terms shown in the square bracket vanishes within the range Eq.~\ref{eq:App_stuRange_IO},
thus, Eq.~\ref{eq:App_InteOvert_Qg_su_tmp} can be simplified as
\begin{equation}
        \begin{aligned}\label{eq:App_InteOvert_Qg_su}
                &\int^{s}_{0}d(-t) ~(-t) \biggr[ \frac{d\sigma}{dt} \biggr]_{Qg(s+u)} \\
		&= \frac{N_{c}^{2}-1}{18\pi\tilde{s}^{\;2}} g^{4} \int^{(-\tilde{u})_{max}}_{(-\tilde{u})_{min}} d(-\tilde{u}) \biggr( -\tilde{u}-\frac{\tilde{u}^{\;2}}{\tilde{s}} + \frac{\tilde{s}^{\;2}}{-\tilde{u}} -\tilde{s} \biggr) \\
		&\approx \frac{N_{c}^{2}-1}{18\pi} g^{4} \biggr( ln\frac{s}{m^2_{1}} - \frac{5}{6} \biggr).
        \end{aligned}
\end{equation}
Note that (1) the boundaries for the integral over $\tilde{u}$ are given in Eq.~\ref{eq:App_uVar_HEA};
(2) the logarithm in Eq.~\ref{eq:App_InteOvert_Qg_su} arises from the term $\int d\tilde{u}/\tilde{u}$ ($u$-channel)
in the range Eq.~\ref{eq:App_stuRange_IO};
(3) in the last step of Eq.~\ref{eq:App_InteOvert_Qg_su} we neglect the terms $\mathcal{O}(\tilde{s}^{-1})$ and $\mathcal{O}(\tilde{s}^{-2})$,
which are suppressed by at least one power of $\tilde{s} \approx s \sim \mathcal{O}(E_{1}T)$ when $E_{1}\rightarrow \infty$.
Inserting Eq.~\ref{eq:App_InteOvert_Qg_su} into Eq.~\ref{eq:App_dEdz_Hard_HEA}, and using Eq.~\ref{eq:App_InteFunc_Bose1} and \ref{eq:App_InteFunc_Bose2}
for the subsequent integral, we can obtain
\begin{equation}
        \begin{aligned}\label{eq:App_dEdx_Hard_HEA_Qg_su}
                &\biggr[-\frac{dE}{dz}\biggr]^{hard-HEA}_{Qg(s+u)} = \frac{N_{c}^{2}-1}{432\pi} g^{4} T^{2} \biggr( ln \frac{4E_{1}T}{m^{2}_{1}}-\frac{5}{6}+c \biggr).
        \end{aligned}
\end{equation}

Combining the contributions from the soft $-t<-t^{\ast}$ (Eq.~\ref{eq:App_dEdx_Soft_HEA})
and hard regions $-t>-t^{\ast}$ (Eq.~\ref{eq:App_dEdx_Hard_HEA_Qq}, \ref{eq:App_dEdx_Hard_HEA_Qg_t} and \ref{eq:App_dEdx_Hard_HEA_Qg_su}),
we find the heavy quark energy loss from scattering off quarks and gluons as
\begin{equation}
	\begin{aligned}
	\biggr[-\frac{dE}{dz}\biggr]^{HEA}_{Qq+Qg} =& \frac{4}{3}\pi\alpha_{s}^{2}T^{2} \biggr[ \bigr( 1 + \frac{N_{f}}{6} \bigr) ln\frac{E_{1}T}{m_{D}^{2}} \\
	& + \frac{2}{9}ln\frac{E_{1}T}{ m^{2}_{1}} + d(N_{f}) \biggr],
	\end{aligned}
\end{equation}
where, the color factor $N_{c}=3$,
the strong coupling factor $g^{2}=4\pi\alpha_{s}$
and the constant parameter $d(N_{f}) \approx 0.146N_{f} + 0.050$.
Similar results can be found in Ref.~\cite{PhysRevD.77.114017}.

 \bibliography{Peng}

\begin{thebibliography}{56}%
\makeatletter
\providecommand \@ifxundefined [1]{%
 \@ifx{#1\undefined}
}%
\providecommand \@ifnum [1]{%
 \ifnum #1\expandafter \@firstoftwo
 \else \expandafter \@secondoftwo
 \fi
}%
\providecommand \@ifx [1]{%
 \ifx #1\expandafter \@firstoftwo
 \else \expandafter \@secondoftwo
 \fi
}%
\providecommand \natexlab [1]{#1}%
\providecommand \enquote  [1]{``#1''}%
\providecommand \bibnamefont  [1]{#1}%
\providecommand \bibfnamefont [1]{#1}%
\providecommand \citenamefont [1]{#1}%
\providecommand \href@noop [0]{\@secondoftwo}%
\providecommand \href [0]{\begingroup \@sanitize@url \@href}%
\providecommand \@href[1]{\@@startlink{#1}\@@href}%
\providecommand \@@href[1]{\endgroup#1\@@endlink}%
\providecommand \@sanitize@url [0]{\catcode `\\12\catcode `\$12\catcode
  `\&12\catcode `\#12\catcode `\^12\catcode `\_12\catcode `\%12\relax}%
\providecommand \@@startlink[1]{}%
\providecommand \@@endlink[0]{}%
\providecommand \url  [0]{\begingroup\@sanitize@url \@url }%
\providecommand \@url [1]{\endgroup\@href {#1}{\urlprefix }}%
\providecommand \urlprefix  [0]{URL }%
\providecommand \Eprint [0]{\href }%
\providecommand \doibase [0]{http://dx.doi.org/}%
\providecommand \selectlanguage [0]{\@gobble}%
\providecommand \bibinfo  [0]{\@secondoftwo}%
\providecommand \bibfield  [0]{\@secondoftwo}%
\providecommand \translation [1]{[#1]}%
\providecommand \BibitemOpen [0]{}%
\providecommand \bibitemStop [0]{}%
\providecommand \bibitemNoStop [0]{.\EOS\space}%
\providecommand \EOS [0]{\spacefactor3000\relax}%
\providecommand \BibitemShut  [1]{\csname bibitem#1\endcsname}%
\let\auto@bib@innerbib\@empty
\bibitem [{\citenamefont {Gyulassy}\ and\ \citenamefont
  {McLerran}(2005)}]{Gyulassy05}%
  \BibitemOpen
  \bibfield  {author} {\bibinfo {author} {\bibfnamefont {M.}~\bibnamefont
  {Gyulassy}}\ and\ \bibinfo {author} {\bibfnamefont {L.}~\bibnamefont
  {McLerran}},\ }\href {\doibase 10.1016/j.nuclphysa.2004.10.034} {\bibfield
  {journal} {\bibinfo  {journal} {Nucl.~Phys.}\ }\textbf {\bibinfo {volume}
  {A750}},\ \bibinfo {pages} {30} (\bibinfo {year} {2005})}\BibitemShut
  {NoStop}%
\bibitem [{\citenamefont {Shuryak}(2005)}]{Shuryak05}%
  \BibitemOpen
  \bibfield  {author} {\bibinfo {author} {\bibfnamefont {E.}~\bibnamefont
  {Shuryak}},\ }\href {\doibase 10.1016/j.nuclphysa.2004.10.022} {\bibfield
  {journal} {\bibinfo  {journal} {Nucl.~Phys.}\ }\textbf {\bibinfo {volume}
  {A750}},\ \bibinfo {pages} {64} (\bibinfo {year} {2005})}\BibitemShut
  {NoStop}%
\bibitem [{\citenamefont {Muller}\ \emph {et~al.}(2012)\citenamefont {Muller},
  \citenamefont {Schukraft},\ and\ \citenamefont {Wyslouch}}]{Muller12}%
  \BibitemOpen
  \bibfield  {author} {\bibinfo {author} {\bibfnamefont {B.}~\bibnamefont
  {Muller}}, \bibinfo {author} {\bibfnamefont {J.}~\bibnamefont {Schukraft}}, \
  and\ \bibinfo {author} {\bibfnamefont {B.}~\bibnamefont {Wyslouch}},\ }\href
  {\doibase 10.1146/annurev-nucl-102711-094910} {\bibfield  {journal} {\bibinfo
   {journal} {Ann.~Rev.~Nucl.~Part.~Sci.}\ }\textbf {\bibinfo {volume} {62}},\
  \bibinfo {pages} {361} (\bibinfo {year} {2012})}\BibitemShut {NoStop}%
\bibitem [{\citenamefont {Shuryak}(2017)}]{Shuryak17}%
  \BibitemOpen
  \bibfield  {author} {\bibinfo {author} {\bibfnamefont {E.}~\bibnamefont
  {Shuryak}},\ }\href {\doibase 10.1103/RevModPhys.89.035001} {\bibfield
  {journal} {\bibinfo  {journal} {Rev.~Mod.~Phys.}\ }\textbf {\bibinfo {volume}
  {89}},\ \bibinfo {pages} {035001} (\bibinfo {year} {2017})}\BibitemShut
  {NoStop}%
\bibitem [{\citenamefont {Kharzeev}\ and\ \citenamefont
  {Liao}(2021)}]{Kharzeev:2020jxw}%
  \BibitemOpen
  \bibfield  {author} {\bibinfo {author} {\bibfnamefont {D.~E.}\ \bibnamefont
  {Kharzeev}}\ and\ \bibinfo {author} {\bibfnamefont {J.}~\bibnamefont
  {Liao}},\ }\href {\doibase 10.1038/s42254-020-00254-6} {\bibfield  {journal}
  {\bibinfo  {journal} {Nature Rev. Phys.}\ }\textbf {\bibinfo {volume} {3}},\
  \bibinfo {pages} {55} (\bibinfo {year} {2021})},\ \Eprint
  {http://arxiv.org/abs/2102.06623} {arXiv:2102.06623 [hep-ph]} \BibitemShut
  {NoStop}%
\bibitem [{\citenamefont {Gyulassy}\ and\ \citenamefont
  {Wang}(1994)}]{GYULASSY1994583}%
  \BibitemOpen
  \bibfield  {author} {\bibinfo {author} {\bibfnamefont {M.}~\bibnamefont
  {Gyulassy}}\ and\ \bibinfo {author} {\bibfnamefont {X.~N.}\ \bibnamefont
  {Wang}},\ }\href {\doibase https://doi.org/10.1016/0550-3213(94)90079-5}
  {\bibfield  {journal} {\bibinfo  {journal} {Nuclear Physics B}\ }\textbf
  {\bibinfo {volume} {420}},\ \bibinfo {pages} {583} (\bibinfo {year}
  {1994})}\BibitemShut {NoStop}%
\bibitem [{\citenamefont {Rapp}\ and\ \citenamefont {van
  Hees}(2010)}]{HQQGPRapp10}%
  \BibitemOpen
  \bibfield  {author} {\bibinfo {author} {\bibfnamefont {R.}~\bibnamefont
  {Rapp}}\ and\ \bibinfo {author} {\bibfnamefont {H.}~\bibnamefont {van
  Hees}},\ }in\ \href {\doibase 10.1142/9789814293297_0003} {\emph {\bibinfo
  {booktitle} {{Quark-Gluon Plasma 4}}}}\ (\bibinfo {year} {2010})\ pp.\
  \bibinfo {pages} {111--206}\BibitemShut {NoStop}%
\bibitem [{\citenamefont {{V.~Greco}}(2017)}]{HFQM17Greco}%
  \BibitemOpen
  \bibfield  {author} {\bibinfo {author} {\bibnamefont {{V.~Greco}}},\ }\href
  {\doibase 10.1016/j.nuclphysa.2017.06.044} {\bibfield  {journal} {\bibinfo
  {journal} {Nucl.~Phys.}\ }\textbf {\bibinfo {volume} {A967}},\ \bibinfo
  {pages} {200} (\bibinfo {year} {2017})}\BibitemShut {NoStop}%
\bibitem [{\citenamefont {{S.~Z.~Shi~{\it et~al}}}(2017)}]{CUJET3QM17}%
  \BibitemOpen
  \bibfield  {author} {\bibinfo {author} {\bibnamefont {{S.~Z.~Shi~{\it
  et~al}}}},\ }\href {\doibase 10.1016/j.nuclphysa.2017.06.037} {\bibfield
  {journal} {\bibinfo  {journal} {Nucl.~Phys.}\ }\textbf {\bibinfo {volume}
  {A967}},\ \bibinfo {pages} {648} (\bibinfo {year} {2017})}\BibitemShut
  {NoStop}%
\bibitem [{\citenamefont {{F.~Prino and R.~Rapp}}(2016)}]{RalfSummary16}%
  \BibitemOpen
  \bibfield  {author} {\bibinfo {author} {\bibnamefont {{F.~Prino and
  R.~Rapp}}},\ }\href {\doibase 10.1088/0954-3899/43/9/093002} {\bibfield
  {journal} {\bibinfo  {journal} {J.~Phys.~G}\ }\textbf {\bibinfo {volume}
  {43}},\ \bibinfo {pages} {093002} (\bibinfo {year} {2016})}\BibitemShut
  {NoStop}%
\bibitem [{\citenamefont {{A.~Andronic~{\it et~al}}}(2016)}]{HFSummaryGROUP16}%
  \BibitemOpen
  \bibfield  {author} {\bibinfo {author} {\bibnamefont {{A.~Andronic~{\it
  et~al}}}},\ }\href {\doibase 10.1140/epjc/s10052-015-3819-5} {\bibfield
  {journal} {\bibinfo  {journal} {Eur.~Phys.~J.}\ }\textbf {\bibinfo {volume}
  {A76}},\ \bibinfo {pages} {107} (\bibinfo {year} {2016})}\BibitemShut
  {NoStop}%
\bibitem [{\citenamefont {{G.~Aarts~{\it et~al}}}(2017)}]{HFSummaryAarts17}%
  \BibitemOpen
  \bibfield  {author} {\bibinfo {author} {\bibnamefont {{G.~Aarts~{\it
  et~al}}}},\ }\href {\doibase 10.1140/epja/i2017-12282-9} {\bibfield
  {journal} {\bibinfo  {journal} {Eur.~Phys.~J.}\ }\textbf {\bibinfo {volume}
  {A54}},\ \bibinfo {pages} {93} (\bibinfo {year} {2017})}\BibitemShut
  {NoStop}%
\bibitem [{\citenamefont {{R.~Rapp~$et.~al.$}}(2018)}]{RalfSummary18}%
  \BibitemOpen
  \bibfield  {author} {\bibinfo {author} {\bibnamefont {{R.~Rapp~$et.~al.$}}},\
  }\href {\doibase 10.1016/j.nuclphysa.2018.09.002} {\bibfield  {journal}
  {\bibinfo  {journal} {Nucl.~Phys.~A}\ }\textbf {\bibinfo {volume} {979}},\
  \bibinfo {pages} {21} (\bibinfo {year} {2018})}\BibitemShut {NoStop}%
\bibitem [{\citenamefont {He}\ \emph {et~al.}(2023)\citenamefont {He},
  \citenamefont {van Hees},\ and\ \citenamefont {Rapp}}]{He:2022ywp}%
  \BibitemOpen
  \bibfield  {author} {\bibinfo {author} {\bibfnamefont {M.}~\bibnamefont
  {He}}, \bibinfo {author} {\bibfnamefont {H.}~\bibnamefont {van Hees}}, \ and\
  \bibinfo {author} {\bibfnamefont {R.}~\bibnamefont {Rapp}},\ }\href {\doibase
  10.1016/j.ppnp.2023.104020} {\bibfield  {journal} {\bibinfo  {journal} {Prog.
  Part. Nucl. Phys.}\ }\textbf {\bibinfo {volume} {130}},\ \bibinfo {pages}
  {104020} (\bibinfo {year} {2023})},\ \Eprint
  {http://arxiv.org/abs/2204.09299} {arXiv:2204.09299 [hep-ph]} \BibitemShut
  {NoStop}%
\bibitem [{\citenamefont {{J.~D.~Bjorken}}(1982)}]{EnergyLossMovivateJDB82}%
  \BibitemOpen
  \bibfield  {author} {\bibinfo {author} {\bibnamefont {{J.~D.~Bjorken}}},\
  }\href@noop {} {\bibfield  {journal} {\bibinfo  {journal}
  {FERMILAB-Pub-82/59-THY}\ } (\bibinfo {year} {1982})}\BibitemShut {NoStop}%
\bibitem [{\citenamefont {Thoma}\ and\ \citenamefont
  {Gyulassy}(1991)}]{THOMA1991491}%
  \BibitemOpen
  \bibfield  {author} {\bibinfo {author} {\bibfnamefont {M.~H.}\ \bibnamefont
  {Thoma}}\ and\ \bibinfo {author} {\bibfnamefont {M.}~\bibnamefont
  {Gyulassy}},\ }\href {\doibase https://doi.org/10.1016/S0550-3213(05)80031-8}
  {\bibfield  {journal} {\bibinfo  {journal} {Nuclear Physics B}\ }\textbf
  {\bibinfo {volume} {351}},\ \bibinfo {pages} {491} (\bibinfo {year}
  {1991})}\BibitemShut {NoStop}%
\bibitem [{\citenamefont {Braaten}\ and\ \citenamefont
  {Thoma}(1991{\natexlab{a}})}]{PhysRevD.44.1298}%
  \BibitemOpen
  \bibfield  {author} {\bibinfo {author} {\bibfnamefont {E.}~\bibnamefont
  {Braaten}}\ and\ \bibinfo {author} {\bibfnamefont {M.~H.}\ \bibnamefont
  {Thoma}},\ }\href {\doibase 10.1103/PhysRevD.44.1298} {\bibfield  {journal}
  {\bibinfo  {journal} {Phys. Rev. D}\ }\textbf {\bibinfo {volume} {44}},\
  \bibinfo {pages} {1298} (\bibinfo {year} {1991}{\natexlab{a}})}\BibitemShut
  {NoStop}%
\bibitem [{\citenamefont {Braaten}\ and\ \citenamefont
  {Thoma}(1991{\natexlab{b}})}]{PhysRevD.44.R2625}%
  \BibitemOpen
  \bibfield  {author} {\bibinfo {author} {\bibfnamefont {E.}~\bibnamefont
  {Braaten}}\ and\ \bibinfo {author} {\bibfnamefont {M.~H.}\ \bibnamefont
  {Thoma}},\ }\href {\doibase 10.1103/PhysRevD.44.R2625} {\bibfield  {journal}
  {\bibinfo  {journal} {Phys. Rev. D}\ }\textbf {\bibinfo {volume} {44}},\
  \bibinfo {pages} {R2625} (\bibinfo {year} {1991}{\natexlab{b}})}\BibitemShut
  {NoStop}%
\bibitem [{\citenamefont {Peign\'e}\ and\ \citenamefont
  {Peshier}(2008{\natexlab{a}})}]{PhysRevD.77.114017}%
  \BibitemOpen
  \bibfield  {author} {\bibinfo {author} {\bibfnamefont {S.}~\bibnamefont
  {Peign\'e}}\ and\ \bibinfo {author} {\bibfnamefont {A.}~\bibnamefont
  {Peshier}},\ }\href {\doibase 10.1103/PhysRevD.77.114017} {\bibfield
  {journal} {\bibinfo  {journal} {Phys. Rev. D}\ }\textbf {\bibinfo {volume}
  {77}},\ \bibinfo {pages} {114017} (\bibinfo {year}
  {2008}{\natexlab{a}})}\BibitemShut {NoStop}%
\bibitem [{\citenamefont {Peshier}(2006)}]{PhysRevLett.97.212301}%
  \BibitemOpen
  \bibfield  {author} {\bibinfo {author} {\bibfnamefont {A.}~\bibnamefont
  {Peshier}},\ }\href {\doibase 10.1103/PhysRevLett.97.212301} {\bibfield
  {journal} {\bibinfo  {journal} {Phys. Rev. Lett.}\ }\textbf {\bibinfo
  {volume} {97}},\ \bibinfo {pages} {212301} (\bibinfo {year}
  {2006})}\BibitemShut {NoStop}%
\bibitem [{\citenamefont {Braaten}\ and\ \citenamefont
  {Pisarski}(1990)}]{PhysRevLett.64.1338}%
  \BibitemOpen
  \bibfield  {author} {\bibinfo {author} {\bibfnamefont {E.}~\bibnamefont
  {Braaten}}\ and\ \bibinfo {author} {\bibfnamefont {R.~D.}\ \bibnamefont
  {Pisarski}},\ }\href {\doibase 10.1103/PhysRevLett.64.1338} {\bibfield
  {journal} {\bibinfo  {journal} {Phys. Rev. Lett.}\ }\textbf {\bibinfo
  {volume} {64}},\ \bibinfo {pages} {1338} (\bibinfo {year}
  {1990})}\BibitemShut {NoStop}%
\bibitem [{\citenamefont {{E.~Braaten and T.~C.~Yuan}}(1991)}]{Braaten91PRL}%
  \BibitemOpen
  \bibfield  {author} {\bibinfo {author} {\bibnamefont {{E.~Braaten and
  T.~C.~Yuan}}},\ }\href {\doibase 10.1103/PhysRevLett.66.2183} {\bibfield
  {journal} {\bibinfo  {journal} {Phys.~Rev.~Lett.}\ }\textbf {\bibinfo
  {volume} {2183}},\ \bibinfo {pages} {66} (\bibinfo {year}
  {1991})}\BibitemShut {NoStop}%
\bibitem [{\citenamefont {Li}\ \emph {et~al.}(2021)\citenamefont {Li},
  \citenamefont {Sun}, \citenamefont {Xie},\ and\ \citenamefont
  {Xiong}}]{Li_2021}%
  \BibitemOpen
  \bibfield  {author} {\bibinfo {author} {\bibfnamefont {S.}~\bibnamefont
  {Li}}, \bibinfo {author} {\bibfnamefont {F.}~\bibnamefont {Sun}}, \bibinfo
  {author} {\bibfnamefont {W.}~\bibnamefont {Xie}}, \ and\ \bibinfo {author}
  {\bibfnamefont {W.}~\bibnamefont {Xiong}},\ }\href {\doibase
  10.1140/epjc/s10052-021-09339-7} {\bibfield  {journal} {\bibinfo  {journal}
  {The European Physical Journal C}\ }\textbf {\bibinfo {volume} {81}}
  (\bibinfo {year} {2021}),\ 10.1140/epjc/s10052-021-09339-7}\BibitemShut
  {NoStop}%
\bibitem [{\citenamefont {Auvinen}\ \emph {et~al.}(2010)\citenamefont
  {Auvinen}, \citenamefont {Eskola},\ and\ \citenamefont
  {Renk}}]{PhysRevC.82.024906}%
  \BibitemOpen
  \bibfield  {author} {\bibinfo {author} {\bibfnamefont {J.}~\bibnamefont
  {Auvinen}}, \bibinfo {author} {\bibfnamefont {K.~J.}\ \bibnamefont {Eskola}},
  \ and\ \bibinfo {author} {\bibfnamefont {T.}~\bibnamefont {Renk}},\ }\href
  {\doibase 10.1103/PhysRevC.82.024906} {\bibfield  {journal} {\bibinfo
  {journal} {Phys. Rev. C}\ }\textbf {\bibinfo {volume} {82}},\ \bibinfo
  {pages} {024906} (\bibinfo {year} {2010})}\BibitemShut {NoStop}%
\bibitem [{\citenamefont {He}\ \emph {et~al.}(2015)\citenamefont {He},
  \citenamefont {Luo}, \citenamefont {Wang},\ and\ \citenamefont
  {Zhu}}]{PhysRevC.91.054908}%
  \BibitemOpen
  \bibfield  {author} {\bibinfo {author} {\bibfnamefont {Y.}~\bibnamefont
  {He}}, \bibinfo {author} {\bibfnamefont {T.}~\bibnamefont {Luo}}, \bibinfo
  {author} {\bibfnamefont {X.-N.}\ \bibnamefont {Wang}}, \ and\ \bibinfo
  {author} {\bibfnamefont {Y.}~\bibnamefont {Zhu}},\ }\href {\doibase
  10.1103/PhysRevC.91.054908} {\bibfield  {journal} {\bibinfo  {journal} {Phys.
  Rev. C}\ }\textbf {\bibinfo {volume} {91}},\ \bibinfo {pages} {054908}
  (\bibinfo {year} {2015})}\BibitemShut {NoStop}%
\bibitem [{\citenamefont {Cao}\ \emph {et~al.}(2016)\citenamefont {Cao},
  \citenamefont {Luo}, \citenamefont {Qin},\ and\ \citenamefont
  {Wang}}]{PhysRevC.94.014909}%
  \BibitemOpen
  \bibfield  {author} {\bibinfo {author} {\bibfnamefont {S.}~\bibnamefont
  {Cao}}, \bibinfo {author} {\bibfnamefont {T.}~\bibnamefont {Luo}}, \bibinfo
  {author} {\bibfnamefont {G.-Y.}\ \bibnamefont {Qin}}, \ and\ \bibinfo
  {author} {\bibfnamefont {X.-N.}\ \bibnamefont {Wang}},\ }\href {\doibase
  10.1103/PhysRevC.94.014909} {\bibfield  {journal} {\bibinfo  {journal} {Phys.
  Rev. C}\ }\textbf {\bibinfo {volume} {94}},\ \bibinfo {pages} {014909}
  (\bibinfo {year} {2016})}\BibitemShut {NoStop}%
\bibitem [{\citenamefont {Caron-Huot}(2009)}]{PhysRevD.79.065039}%
  \BibitemOpen
  \bibfield  {author} {\bibinfo {author} {\bibfnamefont {S.}~\bibnamefont
  {Caron-Huot}},\ }\href {\doibase 10.1103/PhysRevD.79.065039} {\bibfield
  {journal} {\bibinfo  {journal} {Phys. Rev. D}\ }\textbf {\bibinfo {volume}
  {79}},\ \bibinfo {pages} {065039} (\bibinfo {year} {2009})}\BibitemShut
  {NoStop}%
\bibitem [{\citenamefont {{B.~Svetitsky}}(1988)}]{Benjamin88}%
  \BibitemOpen
  \bibfield  {author} {\bibinfo {author} {\bibnamefont {{B.~Svetitsky}}},\
  }\href {\doibase 10.1103/PhysRevD.37.2484} {\bibfield  {journal} {\bibinfo
  {journal} {Phys.~Rev.~D}\ }\textbf {\bibinfo {volume} {37}},\ \bibinfo
  {pages} {2484} (\bibinfo {year} {1988})}\BibitemShut {NoStop}%
\bibitem [{\citenamefont {Weldon}(1982)}]{PhysRevD.26.1394}%
  \BibitemOpen
  \bibfield  {author} {\bibinfo {author} {\bibfnamefont {H.~A.}\ \bibnamefont
  {Weldon}},\ }\href {\doibase 10.1103/PhysRevD.26.1394} {\bibfield  {journal}
  {\bibinfo  {journal} {Phys. Rev. D}\ }\textbf {\bibinfo {volume} {26}},\
  \bibinfo {pages} {1394} (\bibinfo {year} {1982})}\BibitemShut {NoStop}%
\bibitem [{\citenamefont {{P.~B.~Gossiaux and J.~Aichelin}}(2008)}]{PBGPRC08}%
  \BibitemOpen
  \bibfield  {author} {\bibinfo {author} {\bibnamefont {{P.~B.~Gossiaux and
  J.~Aichelin}}},\ }\href {\doibase 10.1103/PhysRevC.78.014904} {\bibfield
  {journal} {\bibinfo  {journal} {Phys.~Rev.~C}\ }\textbf {\bibinfo {volume}
  {78}},\ \bibinfo {pages} {014904} (\bibinfo {year} {2008})}\BibitemShut
  {NoStop}%
\bibitem [{\citenamefont {{J.~P.~Blaizot and E.~Iancu}}(2002)}]{JeanPR02}%
  \BibitemOpen
  \bibfield  {author} {\bibinfo {author} {\bibnamefont {{J.~P.~Blaizot and
  E.~Iancu}}},\ }\href {\doibase 10.1016/S0370-1573(01)00061-8} {\bibfield
  {journal} {\bibinfo  {journal} {Phys.~Rep.}\ }\textbf {\bibinfo {volume}
  {359}},\ \bibinfo {pages} {355} (\bibinfo {year} {2002})}\BibitemShut
  {NoStop}%
\bibitem [{\citenamefont {Romatschke}\ and\ \citenamefont
  {Strickland}(2004)}]{Romatschke:2003vc}%
  \BibitemOpen
  \bibfield  {author} {\bibinfo {author} {\bibfnamefont {P.}~\bibnamefont
  {Romatschke}}\ and\ \bibinfo {author} {\bibfnamefont {M.}~\bibnamefont
  {Strickland}},\ }\href {\doibase 10.1103/PhysRevD.69.065005} {\bibfield
  {journal} {\bibinfo  {journal} {Phys. Rev. D}\ }\textbf {\bibinfo {volume}
  {69}},\ \bibinfo {pages} {065005} (\bibinfo {year} {2004})},\ \Eprint
  {http://arxiv.org/abs/hep-ph/0309093} {arXiv:hep-ph/0309093} \BibitemShut
  {NoStop}%
\bibitem [{\citenamefont {Romatschke}\ and\ \citenamefont
  {Strickland}(2005)}]{Romatschke:2004au}%
  \BibitemOpen
  \bibfield  {author} {\bibinfo {author} {\bibfnamefont {P.}~\bibnamefont
  {Romatschke}}\ and\ \bibinfo {author} {\bibfnamefont {M.}~\bibnamefont
  {Strickland}},\ }\href {\doibase 10.1103/PhysRevD.71.125008} {\bibfield
  {journal} {\bibinfo  {journal} {Phys. Rev. D}\ }\textbf {\bibinfo {volume}
  {71}},\ \bibinfo {pages} {125008} (\bibinfo {year} {2005})},\ \Eprint
  {http://arxiv.org/abs/hep-ph/0408275} {arXiv:hep-ph/0408275} \BibitemShut
  {NoStop}%
\bibitem [{\citenamefont {Djordjevic}(2006)}]{PhysRevC.74.064907}%
  \BibitemOpen
  \bibfield  {author} {\bibinfo {author} {\bibfnamefont {M.}~\bibnamefont
  {Djordjevic}},\ }\href {\doibase 10.1103/PhysRevC.74.064907} {\bibfield
  {journal} {\bibinfo  {journal} {Phys. Rev. C}\ }\textbf {\bibinfo {volume}
  {74}},\ \bibinfo {pages} {064907} (\bibinfo {year} {2006})}\BibitemShut
  {NoStop}%
\bibitem [{\citenamefont {Peign\'e}\ and\ \citenamefont
  {Peshier}(2008{\natexlab{b}})}]{PhysRevD.77.014015}%
  \BibitemOpen
  \bibfield  {author} {\bibinfo {author} {\bibfnamefont {S.}~\bibnamefont
  {Peign\'e}}\ and\ \bibinfo {author} {\bibfnamefont {A.}~\bibnamefont
  {Peshier}},\ }\href {\doibase 10.1103/PhysRevD.77.014015} {\bibfield
  {journal} {\bibinfo  {journal} {Phys. Rev. D}\ }\textbf {\bibinfo {volume}
  {77}},\ \bibinfo {pages} {014015} (\bibinfo {year}
  {2008}{\natexlab{b}})}\BibitemShut {NoStop}%
\bibitem [{\citenamefont {{W.~M.~Alberico, A.~Beraudo, A.~De~Pace, A.~Molinari,
  M.~Monteno, M.~Nardi, and F.~Prino}}(2011)}]{POWLANGEPJC11}%
  \BibitemOpen
  \bibfield  {author} {\bibinfo {author} {\bibnamefont {{W.~M.~Alberico,
  A.~Beraudo, A.~De~Pace, A.~Molinari, M.~Monteno, M.~Nardi, and F.~Prino}}},\
  }\href {\doibase 10.1140/epjc/s10052-011-1666-6} {\bibfield  {journal}
  {\bibinfo  {journal} {Eur.~Phys.~J.~C}\ }\textbf {\bibinfo {volume} {71}},\
  \bibinfo {pages} {1666} (\bibinfo {year} {2011})}\BibitemShut {NoStop}%
\bibitem [{\citenamefont {Carignano}\ and\ \citenamefont
  {Manuel}(2021)}]{PhysRevD.103.116002}%
  \BibitemOpen
  \bibfield  {author} {\bibinfo {author} {\bibfnamefont {S.}~\bibnamefont
  {Carignano}}\ and\ \bibinfo {author} {\bibfnamefont {C.}~\bibnamefont
  {Manuel}},\ }\href {\doibase 10.1103/PhysRevD.103.116002} {\bibfield
  {journal} {\bibinfo  {journal} {Phys. Rev. D}\ }\textbf {\bibinfo {volume}
  {103}},\ \bibinfo {pages} {116002} (\bibinfo {year} {2021})}\BibitemShut
  {NoStop}%
\bibitem [{\citenamefont {{O.~Kaczmarek and F.~Zantow}}(2005)}]{TwoLoopGPRD05}%
  \BibitemOpen
  \bibfield  {author} {\bibinfo {author} {\bibnamefont {{O.~Kaczmarek and
  F.~Zantow}}},\ }\href {\doibase 10.1103/PhysRevD.71.114510} {\bibfield
  {journal} {\bibinfo  {journal} {Phys.~Rev.~D}\ }\textbf {\bibinfo {volume}
  {71}},\ \bibinfo {pages} {114510} (\bibinfo {year} {2005})}\BibitemShut
  {NoStop}%
\bibitem [{\citenamefont {Jeon}\ and\ \citenamefont
  {Moore}(2005)}]{PhysRevC.71.034901}%
  \BibitemOpen
  \bibfield  {author} {\bibinfo {author} {\bibfnamefont {S.}~\bibnamefont
  {Jeon}}\ and\ \bibinfo {author} {\bibfnamefont {G.~D.}\ \bibnamefont
  {Moore}},\ }\href {\doibase 10.1103/PhysRevC.71.034901} {\bibfield  {journal}
  {\bibinfo  {journal} {Phys. Rev. C}\ }\textbf {\bibinfo {volume} {71}},\
  \bibinfo {pages} {034901} (\bibinfo {year} {2005})}\BibitemShut {NoStop}%
\bibitem [{\citenamefont {Qin}\ \emph {et~al.}(2008)\citenamefont {Qin},
  \citenamefont {Ruppert}, \citenamefont {Gale}, \citenamefont {Jeon},
  \citenamefont {Moore},\ and\ \citenamefont
  {Mustafa}}]{PhysRevLett.100.072301}%
  \BibitemOpen
  \bibfield  {author} {\bibinfo {author} {\bibfnamefont {G.-Y.}\ \bibnamefont
  {Qin}}, \bibinfo {author} {\bibfnamefont {J.}~\bibnamefont {Ruppert}},
  \bibinfo {author} {\bibfnamefont {C.}~\bibnamefont {Gale}}, \bibinfo {author}
  {\bibfnamefont {S.}~\bibnamefont {Jeon}}, \bibinfo {author} {\bibfnamefont
  {G.~D.}\ \bibnamefont {Moore}}, \ and\ \bibinfo {author} {\bibfnamefont
  {M.~G.}\ \bibnamefont {Mustafa}},\ }\href {\doibase
  10.1103/PhysRevLett.100.072301} {\bibfield  {journal} {\bibinfo  {journal}
  {Phys. Rev. Lett.}\ }\textbf {\bibinfo {volume} {100}},\ \bibinfo {pages}
  {072301} (\bibinfo {year} {2008})}\BibitemShut {NoStop}%
\bibitem [{\citenamefont {Moore}\ and\ \citenamefont
  {Teaney}(2005)}]{PhysRevC.71.064904}%
  \BibitemOpen
  \bibfield  {author} {\bibinfo {author} {\bibfnamefont {G.~D.}\ \bibnamefont
  {Moore}}\ and\ \bibinfo {author} {\bibfnamefont {D.}~\bibnamefont {Teaney}},\
  }\href {\doibase 10.1103/PhysRevC.71.064904} {\bibfield  {journal} {\bibinfo
  {journal} {Phys. Rev. C}\ }\textbf {\bibinfo {volume} {71}},\ \bibinfo
  {pages} {064904} (\bibinfo {year} {2005})}\BibitemShut {NoStop}%
\bibitem [{\citenamefont {Zhao}\ \emph {et~al.}(2020)\citenamefont {Zhao},
  \citenamefont {Zhou}, \citenamefont {Chen},\ and\ \citenamefont
  {Zhuang}}]{Zhao:2020jqu}%
  \BibitemOpen
  \bibfield  {author} {\bibinfo {author} {\bibfnamefont {J.}~\bibnamefont
  {Zhao}}, \bibinfo {author} {\bibfnamefont {K.}~\bibnamefont {Zhou}}, \bibinfo
  {author} {\bibfnamefont {S.}~\bibnamefont {Chen}}, \ and\ \bibinfo {author}
  {\bibfnamefont {P.}~\bibnamefont {Zhuang}},\ }\href {\doibase
  10.1016/j.ppnp.2020.103801} {\bibfield  {journal} {\bibinfo  {journal} {Prog.
  Part. Nucl. Phys.}\ }\textbf {\bibinfo {volume} {114}},\ \bibinfo {pages}
  {103801} (\bibinfo {year} {2020})},\ \Eprint
  {http://arxiv.org/abs/2005.08277} {arXiv:2005.08277 [nucl-th]} \BibitemShut
  {NoStop}%
\bibitem [{\citenamefont {{Y.~Akamatsu, T.~Hastuda and
  T.~Hirano}}(2009)}]{Akamatsu09}%
  \BibitemOpen
  \bibfield  {author} {\bibinfo {author} {\bibnamefont {{Y.~Akamatsu,
  T.~Hastuda and T.~Hirano}}},\ }\href {\doibase 10.1103/PhysRevC.79.054907}
  {\bibfield  {journal} {\bibinfo  {journal} {Phys.~Rev.~C}\ }\textbf {\bibinfo
  {volume} {75}},\ \bibinfo {pages} {054907} (\bibinfo {year}
  {2009})}\BibitemShut {NoStop}%
\bibitem [{\citenamefont {{M.~He, R.~J.~Fries, and
  R.~Rapp}}(2013)}]{HFModelHee13}%
  \BibitemOpen
  \bibfield  {author} {\bibinfo {author} {\bibnamefont {{M.~He, R.~J.~Fries,
  and R.~Rapp}}},\ }\href {\doibase 10.1103/PhysRevLett.110.112301} {\bibfield
  {journal} {\bibinfo  {journal} {Phys.~Rev.~Lett.}\ }\textbf {\bibinfo
  {volume} {110}},\ \bibinfo {pages} {112301} (\bibinfo {year}
  {2013})}\BibitemShut {NoStop}%
\bibitem [{\citenamefont {{S.~S.~Cao, G.~Y.~Qin, and
  S.~A.~Bass}}(2015)}]{CaoPRC15}%
  \BibitemOpen
  \bibfield  {author} {\bibinfo {author} {\bibnamefont {{S.~S.~Cao, G.~Y.~Qin,
  and S.~A.~Bass}}},\ }\href {\doibase 10.1103/PhysRevC.92.024907} {\bibfield
  {journal} {\bibinfo  {journal} {Phys.~Rev.~C}\ }\textbf {\bibinfo {volume}
  {92}},\ \bibinfo {pages} {024907} (\bibinfo {year} {2015})}\BibitemShut
  {NoStop}%
\bibitem [{\citenamefont {{S.~Li, C.~W.~Wang, X.~B.~Yuan, and
  S.~Q.~Feng}}(2018)}]{CTGUHybrid1}%
  \BibitemOpen
  \bibfield  {author} {\bibinfo {author} {\bibnamefont {{S.~Li, C.~W.~Wang,
  X.~B.~Yuan, and S.~Q.~Feng}}},\ }\href {\doibase 10.1103/PhysRevC.98.014909}
  {\bibfield  {journal} {\bibinfo  {journal} {Phys.~Rev.~C}\ }\textbf {\bibinfo
  {volume} {98}},\ \bibinfo {pages} {014909} (\bibinfo {year}
  {2018})}\BibitemShut {NoStop}%
\bibitem [{\citenamefont {{S.~Li and C.~W.~Wang}}(2018)}]{CTGUHybrid2}%
  \BibitemOpen
  \bibfield  {author} {\bibinfo {author} {\bibnamefont {{S.~Li and
  C.~W.~Wang}}},\ }\href {\doibase 10.1103/PhysRevC.98.034914} {\bibfield
  {journal} {\bibinfo  {journal} {Phys.~Rev.~C}\ }\textbf {\bibinfo {volume}
  {98}},\ \bibinfo {pages} {034914} (\bibinfo {year} {2018})}\BibitemShut
  {NoStop}%
\bibitem [{\citenamefont {{S.~Li, C.~W.~Wang, R.~Z.~Wan, and
  J.~F.~Liao}}(2019)}]{CTGUHybrid3}%
  \BibitemOpen
  \bibfield  {author} {\bibinfo {author} {\bibnamefont {{S.~Li, C.~W.~Wang,
  R.~Z.~Wan, and J.~F.~Liao}}},\ }\href {\doibase 10.1103/PhysRevC.99.054909}
  {\bibfield  {journal} {\bibinfo  {journal} {Phys.~Rev.~C}\ }\textbf {\bibinfo
  {volume} {99}},\ \bibinfo {pages} {054909} (\bibinfo {year}
  {2019})}\BibitemShut {NoStop}%
\bibitem [{\citenamefont {{S. S. Cao, G. Y. Qin, and S. A.
  Bass}}(2013)}]{CaoPRC13}%
  \BibitemOpen
  \bibfield  {author} {\bibinfo {author} {\bibnamefont {{S. S. Cao, G. Y. Qin,
  and S. A. Bass}}},\ }\href {\doibase 10.1103/PhysRevC.88.044907} {\bibfield
  {journal} {\bibinfo  {journal} {Phys.~Rev.~C}\ }\textbf {\bibinfo {volume}
  {88}},\ \bibinfo {pages} {044907} (\bibinfo {year} {2013})}\BibitemShut
  {NoStop}%
\bibitem [{\citenamefont {{S.~Li and J.~F.~Liao}}(2020)}]{CTGUHybrid4}%
  \BibitemOpen
  \bibfield  {author} {\bibinfo {author} {\bibnamefont {{S.~Li and
  J.~F.~Liao}}},\ }\href {\doibase 10.1140/epjc/s10052-020-8243-9} {\bibfield
  {journal} {\bibinfo  {journal} {Eur. Phys. J. C}\ }\textbf {\bibinfo {volume}
  {80}},\ \bibinfo {pages} {671} (\bibinfo {year} {2020})}\BibitemShut
  {NoStop}%
\bibitem [{\citenamefont {Lin}\ \emph {et~al.}(2014)\citenamefont {Lin},
  \citenamefont {Pisarski},\ and\ \citenamefont {Skokov}}]{LIN2014236}%
  \BibitemOpen
  \bibfield  {author} {\bibinfo {author} {\bibfnamefont {S.}~\bibnamefont
  {Lin}}, \bibinfo {author} {\bibfnamefont {R.~D.}\ \bibnamefont {Pisarski}}, \
  and\ \bibinfo {author} {\bibfnamefont {V.~V.}\ \bibnamefont {Skokov}},\
  }\href {\doibase https://doi.org/10.1016/j.physletb.2014.01.043} {\bibfield
  {journal} {\bibinfo  {journal} {Physics Letters B}\ }\textbf {\bibinfo
  {volume} {730}},\ \bibinfo {pages} {236} (\bibinfo {year}
  {2014})}\BibitemShut {NoStop}%
\bibitem [{\citenamefont {{S.~Li, W.~Xiong, and
  R.~Z.~Wan}}(2020)}]{CTGUHybrid5}%
  \BibitemOpen
  \bibfield  {author} {\bibinfo {author} {\bibnamefont {{S.~Li, W.~Xiong, and
  R.~Z.~Wan}}},\ }\href {\doibase 10.1140/epjc/s10052-020-08708-y} {\bibfield
  {journal} {\bibinfo  {journal} {Eur. Phys. J. C}\ }\textbf {\bibinfo {volume}
  {80}},\ \bibinfo {pages} {1113} (\bibinfo {year} {2020})}\BibitemShut
  {NoStop}%
\bibitem [{\citenamefont {Das}\ \emph {et~al.}(2014)\citenamefont {Das},
  \citenamefont {Scardina}, \citenamefont {Plumari},\ and\ \citenamefont
  {Greco}}]{PhysRevC.90.044901}%
  \BibitemOpen
  \bibfield  {author} {\bibinfo {author} {\bibfnamefont {S.~K.}\ \bibnamefont
  {Das}}, \bibinfo {author} {\bibfnamefont {F.}~\bibnamefont {Scardina}},
  \bibinfo {author} {\bibfnamefont {S.}~\bibnamefont {Plumari}}, \ and\
  \bibinfo {author} {\bibfnamefont {V.}~\bibnamefont {Greco}},\ }\href
  {\doibase 10.1103/PhysRevC.90.044901} {\bibfield  {journal} {\bibinfo
  {journal} {Phys. Rev. C}\ }\textbf {\bibinfo {volume} {90}},\ \bibinfo
  {pages} {044901} (\bibinfo {year} {2014})}\BibitemShut {NoStop}%
\bibitem [{\citenamefont {Xu}\ \emph {et~al.}(2019)\citenamefont {Xu},
  \citenamefont {Bass}, \citenamefont {Moreau}, \citenamefont {Song},
  \citenamefont {Nahrgang}, \citenamefont {Bratkovskaya}, \citenamefont
  {Gossiaux}, \citenamefont {Aichelin}, \citenamefont {Cao}, \citenamefont
  {Greco}, \citenamefont {Coci},\ and\ \citenamefont
  {Werner}}]{XuCoefficient18}%
  \BibitemOpen
  \bibfield  {author} {\bibinfo {author} {\bibfnamefont {Y.}~\bibnamefont
  {Xu}}, \bibinfo {author} {\bibfnamefont {S.~A.}\ \bibnamefont {Bass}},
  \bibinfo {author} {\bibfnamefont {P.}~\bibnamefont {Moreau}}, \bibinfo
  {author} {\bibfnamefont {T.}~\bibnamefont {Song}}, \bibinfo {author}
  {\bibfnamefont {M.}~\bibnamefont {Nahrgang}}, \bibinfo {author}
  {\bibfnamefont {E.}~\bibnamefont {Bratkovskaya}}, \bibinfo {author}
  {\bibfnamefont {P.}~\bibnamefont {Gossiaux}}, \bibinfo {author}
  {\bibfnamefont {J.}~\bibnamefont {Aichelin}}, \bibinfo {author}
  {\bibfnamefont {S.}~\bibnamefont {Cao}}, \bibinfo {author} {\bibfnamefont
  {V.}~\bibnamefont {Greco}}, \bibinfo {author} {\bibfnamefont
  {G.}~\bibnamefont {Coci}}, \ and\ \bibinfo {author} {\bibfnamefont
  {K.}~\bibnamefont {Werner}},\ }\href {\doibase 10.1103/PhysRevC.99.014902}
  {\bibfield  {journal} {\bibinfo  {journal} {Phys. Rev. C}\ }\textbf {\bibinfo
  {volume} {99}},\ \bibinfo {pages} {014902} (\bibinfo {year}
  {2019})}\BibitemShut {NoStop}%
\bibitem [{\citenamefont {Gubser}(2008)}]{GUBSER2008175}%
  \BibitemOpen
  \bibfield  {author} {\bibinfo {author} {\bibfnamefont {S.~S.}\ \bibnamefont
  {Gubser}},\ }\href {\doibase https://doi.org/10.1016/j.nuclphysb.2007.09.017}
  {\bibfield  {journal} {\bibinfo  {journal} {Nuclear Physics B}\ }\textbf
  {\bibinfo {volume} {790}},\ \bibinfo {pages} {175} (\bibinfo {year}
  {2008})}\BibitemShut {NoStop}%
\bibitem [{\citenamefont {{B.~L.~Combridge}}(1979)}]{Combridge79}%
  \BibitemOpen
  \bibfield  {author} {\bibinfo {author} {\bibnamefont {{B.~L.~Combridge}}},\
  }\href {\doibase 10.1016/0550-3213(79)90449-8} {\bibfield  {journal}
  {\bibinfo  {journal} {Nucl.~Phys.~B}\ }\textbf {\bibinfo {volume} {151}},\
  \bibinfo {pages} {429} (\bibinfo {year} {1979})}\BibitemShut {NoStop}%
\end{thebibliography}%
\end{document}